\documentclass[conference]{IEEEtran}
\usepackage{microtype} 
\usepackage{graphicx}
\usepackage[hyphens]{url}
\usepackage{booktabs}
\usepackage[tight,footnotesize]{subfigure}
\usepackage{algorithm}
\usepackage{algpseudocode}
\usepackage{eurosym}
\usepackage{hyperref}
\usepackage{xspace}
\usepackage[T1]{fontenc}
\usepackage[table]{xcolor}
\usepackage{balance}

\usepackage{tabularx}

\usepackage{pifont}               
\usepackage{textcomp}             
\usepackage[cmex10]{amsmath}      
\usepackage{amssymb}
\usepackage{listings}
\hypersetup{%
  pdftitle={Thou Shalt Not Depend on Me: Analysing the Use of Outdated JavaScript Libraries on the Web},
  pdfauthor={Tobias Lauinger, Abdelberi Chaabane, William Robertson, Christo Wilson, Engin Kirda},
  pdfkeywords={{Web Security}, {JavaScript Libraries}, {Vulnerabilities}, {Web Site Maintenance}},
  pdfsubject={},
  hidelinks=true, 
  colorlinks=false,   
  linkcolor=black,
  citecolor=black,
  urlcolor=black,
  pdfstartview=FitB} 

\usepackage[all]{hypcap} 

\usepackage[firstpage=true,scale=1,angle=0,opacity=1,color=red,position={current page.north west},nodeanchor={north west},hshift=0.680in,vshift=-0.18in]{background}
\SetBgContents{\parbox{\textwidth}{\footnotesize
\textbf{Updated in September 2017:}
Require valid versions for library detection throughout the paper.
The vulnerability analysis already did so and remains identical.
Modifications in
Tables~\ref{tab:libraries}, \ref{tab:hostRankingLibraries}
and~\ref{tab:libraryInclusions}; Figures~\ref{fig:ditrib_detected_alexa}
and~\ref{fig:ditrib_detected_com}; Sections~\ref{sec:identification},
\ref{sec:analysis:general}, \ref{sec:analysis:detections},
\ref{sec:analysis:age} and~\ref{sec:analysis:aliasing}.
Additionally, highlight Ember's security practices in
Section~\ref{sec:discussion}.%
}}

\newcommand{\para}[1]{{\vspace{4pt} \bf \noindent #1 \hspace{10pt}}}

\newenvironment{packed_itemize}{
\begin{itemize}
  \setlength{\itemsep}{2pt}
  \setlength{\parskip}{0pt}
  \setlength{\parsep}{0pt}
  \setlength{\topsep}{2pt}
}{\end{itemize}}

\newcommand{\eg}{e.g.,\ }
\newcommand{\etal}{et al.\xspace}
\newcommand{\ie}{i.e.,\ }

\begin{document}
\bstctlcite{IEEEexample:BSTcontrol}

\title{Thou Shalt Not Depend on Me: Analysing the Use of Outdated JavaScript Libraries on the Web}

\author{\IEEEauthorblockN{Tobias Lauinger, Abdelberi Chaabane,
Sajjad Arshad, William Robertson, Christo Wilson and Engin Kirda}
\IEEEauthorblockA{Northeastern University\\
\{toby, 3abdou, arshad, wkr, cbw, ek\}@ccs.neu.edu}}

\IEEEoverridecommandlockouts
\makeatletter\def\@IEEEpubidpullup{9\baselineskip}\makeatother
\IEEEpubid{\parbox{\columnwidth}{Permission to freely reproduce all or part
    of this paper for noncommercial purposes is granted provided that
    copies bear this notice and the full citation on the first
    page. Reproduction for commercial purposes is strictly prohibited
    without the prior written consent of the Internet Society, the
    first-named author (for reproduction of an entire paper only), and
    the author's employer if the paper was prepared within the scope
    of employment.  \\
    NDSS '17, 26 February - 1 March 2017, San Diego, CA, USA\\
    Copyright 2017 Internet Society, ISBN 1-1891562-46-0\\
    https://doi.org/10.14722/ndss.2017.23414
}
\hspace{\columnsep}\makebox[\columnwidth]{}}

\maketitle

\begin{abstract}
Web developers routinely rely on third-party Java\-Script libraries such as
jQuery to enhance the functionality of their sites. However, if not
properly maintained, such dependencies can create attack vectors allowing
a site to be compromised.

In this paper, we conduct the first comprehensive study of client-side
JavaScript library usage and the resulting security implications across
the Web.
Using data from over 133\,k websites, we show that 37\,\% of them include
at least one library with a known vulnerability; the time lag behind the
newest release of a library is measured in the order of years.
In order to better understand why websites use so many vulnerable or
outdated libraries, we track causal inclusion relationships and quantify
different scenarios.
We observe sites including libraries in ad hoc and often transitive ways,
which can lead to different versions of the same library being loaded
into the same document at the same time.
Furthermore, we find that libraries included transitively, or via ad and
tracking code, are more likely to be vulnerable. This demonstrates that
not only website administrators, but also the dynamic architecture and
developers of third-party services are to blame for the Web's poor
state of library management.

The results of our work underline the need for more thorough approaches to
dependency management, code maintenance and third-party code inclusion on
the Web.
\end{abstract}

\section{Introduction}

The Web is arguably the most popular contemporary programming platform.
Although websites are relatively easy to create, they are often composed of
heterogeneous components such as database backends, content generation engines, multiple
scripting languages and client-side code, and they need to deal with unsanitised inputs
encoded in several different formats. Hence, it is no surprise that it is
challenging to secure websites because of the large attack surface they
expose.

One specific, significant attack surface are vulnerabilities related to
client-side JavaScript, such as cross-site scripting (XSS) and advanced phishing.
Crucially, modern websites often include popular third-party JavaScript libraries,
and thus are at risk of inheriting vulnerabilities contained in these libraries.
For example,
a 2013 XSS vulnerability in the jQuery~\cite{jquery} library before version 1.6.3
allowed remote attackers to inject arbitrary scripts or HTML into vulnerable
websites via a crafted tag. As a result, it is of
the utmost importance for websites to manage library dependencies and,
in particular, to update vulnerable libraries in a timely fashion.

To date, security research has addressed a wide range of client-side security
issues in websites, including validation~\cite{saxena10flax} and
XSS~(\cite{johns13ccs,vogt07}),
cross-site request forgery~\cite{barth08}, and session fixation~\cite{takamatsu10}.
However, the use of vulnerable JavaScript libraries by websites has not
received nearly as much attention. In 2014, a series of blog posts presented
cursory measurements highlighting that major websites included known vulnerable
libraries~(\cite{blog2,blog3,blog1}). These findings echo warnings from other
software ecosystems like Android~\cite{backes16}, Java~\cite{sonatype}
and Windows~\cite{nappa15}, which show that vulnerable libraries continue
to exist in the wild even when
they are widely known to contain severe vulnerabilities. Given that JavaScript
dependency management is relatively primitive and corresponding tools are
not as well-established as in more mature ecosystems, these findings
suggest that security issues caused by
outdated JavaScript libraries on the Web may be widespread.

In this paper, we conduct the first comprehensive study on the security
implications of JavaScript library usage in websites. We seek to answer
the following questions:
\begin{packed_itemize}

\item Where do websites load JavaScript libraries from (\ie first
or third-party domains), and how frequently are these domains used?

\item How current are the libraries that websites are using, and do they
contain known vulnerabilities?

\item Are web developers intentionally including JavaScript libraries, or
are these dependencies caused by advertising and tracking code?

\item Are existing remediation strategies effective or widely used?

\item Are there additional technical, methodological, or organisational
changes that can
improve the security of websites with respect to JavaScript library usage?

\end{packed_itemize}
Note that the focus of this paper is {\em not} measuring the security state of
specific JavaScript libraries.
Rather, our goal (and primary contribution) is to empirically examine whether website
operators keep their libraries current and react to publicly disclosed vulnerabilities.

Answering these questions necessitated solving three fundamental methodological
challenges. {\em First}, there is no centralised repository of metadata
pertaining to JavaScript libraries and their versions, release dates, and
known vulnerabilities. To address this, we manually constructed a catalogue
containing all ``release'' versions of 72 of the most popular open-source
libraries, including detailed vulnerability information on a subset of 11
libraries.  {\em Second}, web developers often modify JavaScript libraries
by reformatting, restructuring or appending code, which makes it difficult
to detect library usage in the wild. We solve this problem through a
combination of
static and dynamic analysis techniques. {\em Third}, to understand why specific
libraries are loaded by a given site, we need to track all of the causal
relationships between page elements (\eg script $s_1$ in frame $f_1$
injects script $s_2$ into frame $f_2$). To solve this, we developed a
customised version of Chromium that records detailed {\em causality trees}
of page element creation relationships.

Using these tools, we crawled the Alexa Top~75\,k websites and a random
sample of 75\,k websites drawn from a snapshot of the \texttt{.com} zone
in May~2016. These
two crawls allow us to compare and contrast JavaScript library usage between popular
and unpopular websites. In total, we observed 11,141,726 inline scripts
and script file inclusions; 86.6\,\% of Alexa sites and 65.4\,\% of
\texttt{.com} sites used at least one well-known
JavaScript library, with jQuery being the most popular by a large majority.

Analysis of our dataset reveals many concerning facts about JavaScript library
management on today's Web. More than a third of the websites in our Alexa
crawl include at least one vulnerable library version, and nearly 10\,\%
include two or more different vulnerable versions. From a per-library
perspective, at least 36.7\,\% of jQuery, 40.1\,\% of Angular, 86.6\,\%
of Handlebars, and 87.3\,\% of YUI inclusions use a vulnerable version.
Alarmingly, many sites continue to rely on libraries like YUI and SWFObject that
are no longer maintained. In fact, the median website in our dataset is
using a library version 1,177~days older than the newest release,
which explains why so many vulnerable libraries tend to linger
on the Web. Advertising, tracking and social widget code 
can cause \textit{transitive} library inclusions with
a higher rate of vulnerability, suggesting that
these problems extend beyond individual website administrators to providers
of Web infrastructure and services.

We also observe many websites exhibiting surprising behaviours with
respect to JavaScript library inclusion. For example, 4.2\,\% of websites
using jQuery in
the Alexa crawl include the same library version multiple times in the same
document, and 10.9\,\% include multiple \textit{different versions} of
jQuery into the \textit{same document}. To our knowledge, ours is the
first study to make these observations, since existing
tools~(\cite{retirejs,librarydetector}) are unable to detect these
anomalies. These strange behaviours may have a negative impact on security
as asynchronous loading leads to nondeterministic behaviour, and it
remains unclear which version will ultimately be used.

Perhaps our most sobering finding is practical evidence that the
JavaScript library ecosystem is complex, unorganised, and quite
``ad hoc'' with respect to security. As of this writing, there are no
reliable vulnerability databases, no security mailing lists for the most
popular libraries, few
or no details on security issues in release notes, and often,
it is difficult to identify which versions of a library are affected by a
reported vulnerability.

Overall, our study makes the following contributions:
\begin{packed_itemize}

\item We conduct the first comprehensive study showing that a significant number of
websites include vulnerable or outdated JavaScript libraries.

\item We present results on the origins of vulnerable JavaScript library
inclusions, which allows us to contrast the security posture of website
developers with third-party modules such as WordPress, advertising or
tracking networks, and social media widgets.

\item We show that a large number of websites include JavaScript libraries in
unexpected ways, such as multiple inclusions of different library versions
into the same document, which may impact their attack surface.

\item We find existing remediation strategies to be ineffective at
mitigating the threats posed by vulnerable JavaScript libraries. For
example, less than 3\,\% of websites could fix all their vulnerable
libraries by applying only patch-level updates. Similarly, only 2\,\% of
websites use the
version-aliasing services offered by JavaScript CDNs.

\end{packed_itemize}

\section{Background}
\label{sec:background}
JavaScript has allowed web developers to build highly interactive websites
with sophisticated functionality. For example, communication and
production-related online services such as
Gmail and Office~365 make heavy use of JavaScript to create web-based
applications comparable to their more traditional desktop counterparts.
In this paper, we focus exclusively on aspects of client-side JavaScript
executed in a browser, not the recent trend of using JavaScript for
server-side programming.

\subsection{JavaScript Libraries}

In many cases,
to make their lives easier, web developers rely on functionality that is bundled
in libraries. For example, jQuery~\cite{jquery} is a popular JavaScript library
that makes HTML
document traversal and manipulation, event handling, animation, and AJAX much
simpler and compatible across browsers.

In the simplest case, a JavaScript library is a plain-text script
containing code with reasonably well-defined functionality.  The script has
full access to the DOM that includes it; the
concept of namespaces does not exist in JavaScript, and everything that is
created is by default {\em global}.  More elaborate libraries use hacks and
conventions to protect the code against naming conflicts, and expose interfaces
for retrieving meta-data such as the name and version of the library. Over the
course of this study, we found that JavaScript libraries overwhelmingly
use the Semantic Versioning~\cite{semanticVersioning} convention of
\texttt{major.minor.patch}, such as \texttt{1.0.1}, where the major
version component is increased for breaking changes, the minor component for new
functionality, and the patch component for backwards-compatible bug fixes.

To include a library into their website, developers typically use the
\texttt{<script src="url"></script>} HTML tag and point to
an externally-hosted version of the library or a copy on
their own server. Library vendors often provide a \textit{minified}
version that has comments and whitespace removed and local variables
shortened to reduce the size of the file.
Developers can also concatenate multiple libraries into
a single file, create custom builds of libraries, or use advanced
minification features such as dead code removal. While custom minification
builds are relatively common, more aggressive minification settings are
rare in client-side JavaScript because they can break
code~\cite{closureCompilerMinificationSettings}.

\para{CDNs.} Many libraries are available on Content Distribution
Networks (CDNs) for use by other websites. Google, Microsoft and Yandex
host libraries on their CDNs, some popular libraries (\eg Bootstrap and
jQuery) offer their own CDNs, and some community-based CDNs
accept to host arbitrary open-source libraries. JavaScript CDNs
enable caching of libraries across websites to increase performance.
Another useful feature offered by some CDNs is {\em version aliasing}.
That is, when including a library, the
developer may specify a version prefix instead of the full version string,
in which case the CDN returns the
newest available version with that prefix. When implemented correctly, the patched
version of a library will automatically be used on the website when it becomes
available on the CDN. However, this works only for security issues
fixed in a backwards-compatible manner, and it conflicts with client-side
security mechanisms such as subresource integrity~\cite{subresource_integrity}.
In addition, version aliasing makes client-side caching of resources
less efficient because it must be configured for shorter time
spans, that is, hours instead of years. As a result, version aliasing
is often discouraged~\cite{googleCDN}.

\para{Third Parties.} Third-party modules such as advertising, trackers,
social media or other widgets that are often embedded in webpages
typically implemented in JavaScript.
Furthermore, these scripts can also load libraries, possibly without the
knowledge of the site maintainer. If not isolated in a frame, these
libraries gain full privileges in the including site's context. Thus, even if a
web developer keeps own library dependencies updated, outdated versions may still be
included by badly maintained third-party content. Also, some JavaScript
libraries and many web frameworks contain their own copies of libraries they
depend on. Hence, web developers may unknowingly rely on software maintainers
to update JavaScript libraries.

\subsection{Vulnerabilities in JavaScript Libraries}

While JavaScript is the de-facto standard for developing client-side code on
the Web, at the same time it is notorious for security vulnerabilities.
A common, lingering problem
is Cross-Site Scripting (XSS)~\cite{johns13ccs}, which allows an attacker to
inject malicious code (or HTML) into a website.  In particular, if a JavaScript
library accepts input from the user and does not validate it, an XSS
vulnerability might creep in, and all websites using this library could
become vulnerable.

As an example, consider the popular jQuery library and its \texttt{\$()}
function, which is overloaded and has different behaviour depending on
which type of argument is passed~\cite{jQueryDocumentation}: If a string
containing a CSS selector is passed, the function searches the DOM tree
for corresponding elements and returns references to them; if the input
string contains HTML, the function creates the corresponding elements and
returns the references. As a consequence, developers who pass improperly
sanitised input to this function may inadvertently allow attackers to
inject code even though the developers' original intent was only to
select an existing element. While this API design places convenience over
security considerations and the implications could be better highlighted
in the documentation, it does not automatically constitute a
vulnerability in the library.

In older versions of jQuery, however, the \texttt{\$()} function's
leniency in parsing string parameters could lead to complications by
misleading developers to believe, for instance, that any string beginning
with \texttt{\#} would be interpreted as a selector and could be safe to
pass to the function, as \texttt{\#test} selects the element with
identifier \texttt{test}. Yet, jQuery considered parameters containing a
HTML \texttt{<tag>} \textit{anywhere} in the string as
HTML~\cite{jQueryBug}, so that a parameter such as
\texttt{\#<img src=/ onerror=alert(1)>} would lead to code execution
rather than a selection. This behaviour was considered a
vulnerability and fixed.

Other vulnerabilities in JavaScript libraries include cases
where libraries do not sanitise inputs that are expected to be pure
text, but are passed to \texttt{eval()} or \texttt{document.write()}
internally, which could cause them to be executed as
script or rendered as markup, respectively. Attackers can use
these capabilities to steal data from a user's browsing session, initiate
transactions on the user's behalf, or place fake content on a website.
Therefore, JavaScript libraries must not
introduce any attack vectors into the websites where they are used.

\begin{table}[t]
\caption{The 30 most frequent libraries in our \textsc{Alexa}
crawl (out of 72 supported libraries).
Versions: \textcircled{s} total in reference catalogue for static
detection;
\textcircled{d} dynamic detections observed in crawls (some
libraries/versions not supported dynamically).}
\centering 
\scriptsize
\rowcolors{2}{gray!10}{white}

\begin{tabular}{lrr|rr|rr}
\toprule
 & \multicolumn{2}{c|}{\textbf{Versions}} & \textbf{Bower} & \textbf{Wapp} & \multicolumn{2}{c}{\textbf{Use on Crawled Sites}} \\
\textbf{Library} & \textcircled{s} & \textcircled{d} & \textbf{Rank} & \textbf{\%} & \textbf{\textsc{Alexa}} & \textbf{\textsc{Com}} \\
\midrule
jQuery & 66 & 64 & 1 & 42\,\% & \textcolor{black}{83.9\,\%} & \textcolor{black}{61.1\,\%} \\
jQuery-UI & 46 & 46 & 13 & 7\,\% & \textcolor{black}{23.5\,\%} & \textcolor{black}{8.0\,\%} \\
Modernizr & 24 & 28 & 18 & 10\,\% & \textcolor{black}{21.4\,\%} & \textcolor{black}{8.6\,\%} \\
Bootstrap & 32 & 10 & 3 &  & \textcolor{black}{12.5\,\%} & \textcolor{black}{4.5\,\%} \\
jQuery-Migrate & 7 & 0 &  &  & \textcolor{black}{11.3\,\%} & \textcolor{black}{10.7\,\%} \\
Underscore & 61 & 34 & 12 & 3\,\% & \textcolor{black}{5.8\,\%} & \textcolor{black}{2.4\,\%} \\
SWFObject & 2 & 1 &  &  & \textcolor{black}{3.7\,\%} & \textcolor{black}{2.4\,\%} \\
Moment & 54 & 33 & 6 &  & \textcolor{black}{3.5\,\%} & \textcolor{black}{1.4\,\%} \\
RequireJS & 62 & 40 &  &  & \textcolor{black}{3.4\,\%} & \textcolor{black}{2.3\,\%} \\
jQuery-Form & 14 & 0 &  &  & \textcolor{black}{2.7\,\%} & \textcolor{black}{3.4\,\%} \\
Backbone & 29 & 19 &  & 2\,\% & \textcolor{black}{2.7\,\%} & \textcolor{black}{1.6\,\%} \\
Angular & 110 & 78 & 2 &  & \textcolor{black}{2.4\,\%} & \textcolor{black}{1.6\,\%} \\
LoDash & 77 & 57 & 26 &  & \textcolor{black}{2.4\,\%} & \textcolor{black}{2.5\,\%} \\
jQuery-Tools & 8 & 20 &  &  & \textcolor{black}{2.3\,\%} & \textcolor{black}{0.9\,\%} \\
jQuery-Fancybox & 10 & 0 &  &  & \textcolor{black}{2.3\,\%} & \textcolor{black}{1.4\,\%} \\
GreenSock GSAP & 45 & 45 &  &  & \textcolor{black}{2.2\,\%} & \textcolor{black}{2.8\,\%} \\
Handlebars & 25 & 15 &  &  & \textcolor{black}{2.0\,\%} & \textcolor{black}{0.4\,\%} \\
Prototype & 5 & 14 &  &  & \textcolor{black}{1.8\,\%} & \textcolor{black}{1.1\,\%} \\
MooTools & 27 & 24 &  & 3\,\% & \textcolor{black}{1.5\,\%} & \textcolor{black}{1.4\,\%} \\
WebFont Loader & 100 & 0 &  &  & \textcolor{black}{1.5\,\%} & \textcolor{black}{0.9\,\%} \\
jQuery-Cookie & 8 & 0 & 21 &  & \textcolor{black}{1.4\,\%} & \textcolor{black}{0.2\,\%} \\
Hammer.js & 26 & 14 &  &  & \textcolor{black}{1.2\,\%} & \textcolor{black}{0.4\,\%} \\
jQuery-Validation & 13 & 0 &  &  & \textcolor{black}{1.1\,\%} & \textcolor{black}{0.6\,\%} \\
Mustache & 29 & 21 &  &  & \textcolor{black}{1.1\,\%} & \textcolor{black}{0.9\,\%} \\
YUI 3 & 37 & 26 &  &  & \textcolor{black}{1.0\,\%} & \textcolor{black}{1.4\,\%} \\
Velocity & 55 & 15 &  &  & \textcolor{black}{0.9\,\%} & \textcolor{black}{0.2\,\%} \\
Script.aculo.us & 5 & 12 &  &  & \textcolor{black}{0.8\,\%} & \textcolor{black}{0.4\,\%} \\
Knockout & 21 & 9 &  &  & \textcolor{black}{0.8\,\%} & \textcolor{gray}{0.1\,\%} \\
Flexslider & 11 & 0 &  &  & \textcolor{black}{0.6\,\%} & \textcolor{black}{0.4\,\%} \\
React & 41 & 23 & 28 &  & \textcolor{black}{0.5\,\%} & \textcolor{black}{1.6\,\%} \\
\bottomrule
\end{tabular}

\label{tab:libraries}
\end{table}

\section{Methodology}
\label{sec:methodology}

Identifying client-side JavaScript libraries, finding out how they
are loaded by a website, and determining whether they are outdated or
vulnerable requires a combination of techniques and data sources.
Challenges arise due to the
lax JavaScript language, the fragmented
library ecosystem, and the complex nature of modern websites.
{\em First}, we need to collect metadata about popular JavaScript
libraries, including a list of available versions, the corresponding
release dates, code samples, and known vulnerabilities. {\em Second}, we
must be able to determine if JavaScript code found in the wild is a known
library. {\em Third}, we need to crawl websites while keeping
track of causal resource inclusion relationships and match them with
detected libraries.

\subsection{Catalogueing JavaScript Libraries}

In contrast to Maven's Central Repository in the Java world,
JavaScript does not have a similarly popular
repository of library versioning and project dependency metadata. We must
therefore collect and correlate this data from various separate
sources.

\subsubsection{Selecting Libraries}

The initial
construction of our metadata archive involves a certain amount of manual
verification work. Since there are thousands of JavaScript libraries (\eg
the community-based \url{cdnjs.com} hosts 2,379 projects as of
August~2016), we focus our study on the most widely used libraries
because they are the most consequential.

To select libraries, we leverage
library popularity statistics provided by the JavaScript package manager
Bower~\cite{bower} and the web technology survey
Wappalyzer~\cite{wappalyzer}.
We extend this list of popular libraries with all projects hosted on the
public CDNs operated by Google, Microsoft and Yandex. As we will show in
Section~\ref{sec:analysis:detections}, many websites rely on these commercial
CDNs to host JavaScript libraries. We collected the data from Bower,
Wappalyzer, and the three CDNs in January 2016.

Due to various data availability requirements explained in detail
in Section~\ref{sec:libcatalogue:limits}, we need to exclude certain libraries
from our study. Overall, we support 72~libraries---18~out of the Top~20
installed with Bower, 7~out of the Top~10 frameworks identified on
websites by Wappalyzer, 13~of the 14~libraries hosted by Google, 12~of
the 18~libraries hosted by Microsoft, and all 11 libraries hosted by
Yandex. Table~\ref{tab:libraries} shows a subset of 30 libraries in our
catalogue as well as their rank on Bower and their market share according
to Wappalyzer. Although our catalogue appears to cover a sparse set of
the libraries on Bower, many of the missing ranks belong to submodules
of popular libraries (\eg rank~5 is {\em Angular Mocks}). According to
Wappalyzer, we cover 73\,\% of the most popular libraries.

\subsubsection{Extracting Versioning Information}

Our next step is compiling a
complete list of library versions along with their release dates. After
unsuccessful experiments with file timestamps and available-since dates
on the libraries' official websites and CDNs, we determined that GitHub
was the most reliable source for this kind of information. Nearly all of
the open source libraries in our seed lists are hosted on GitHub
and tag the source code of their releases, allowing us to extract
timestamps and version identifiers from the tags. In naming their
releases, they typically follow a \texttt{major.minor.patch} version
numbering scheme, which makes it straightforward to identify tags
pertaining to releases and ignore all other tags, including ``alpha,''
``beta'' and ``release candidate'' versions that are not meant to be used
in production. As shown in Table~\ref{tab:libraries}, popular libraries
like Angular and jQuery have up to 110 and 66 distinct versions in our
catalogue, respectively. However, half of the libraries have fewer than
26~versions.

\begin{figure}[t]
\centering
\includegraphics[width=1.0\columnwidth]{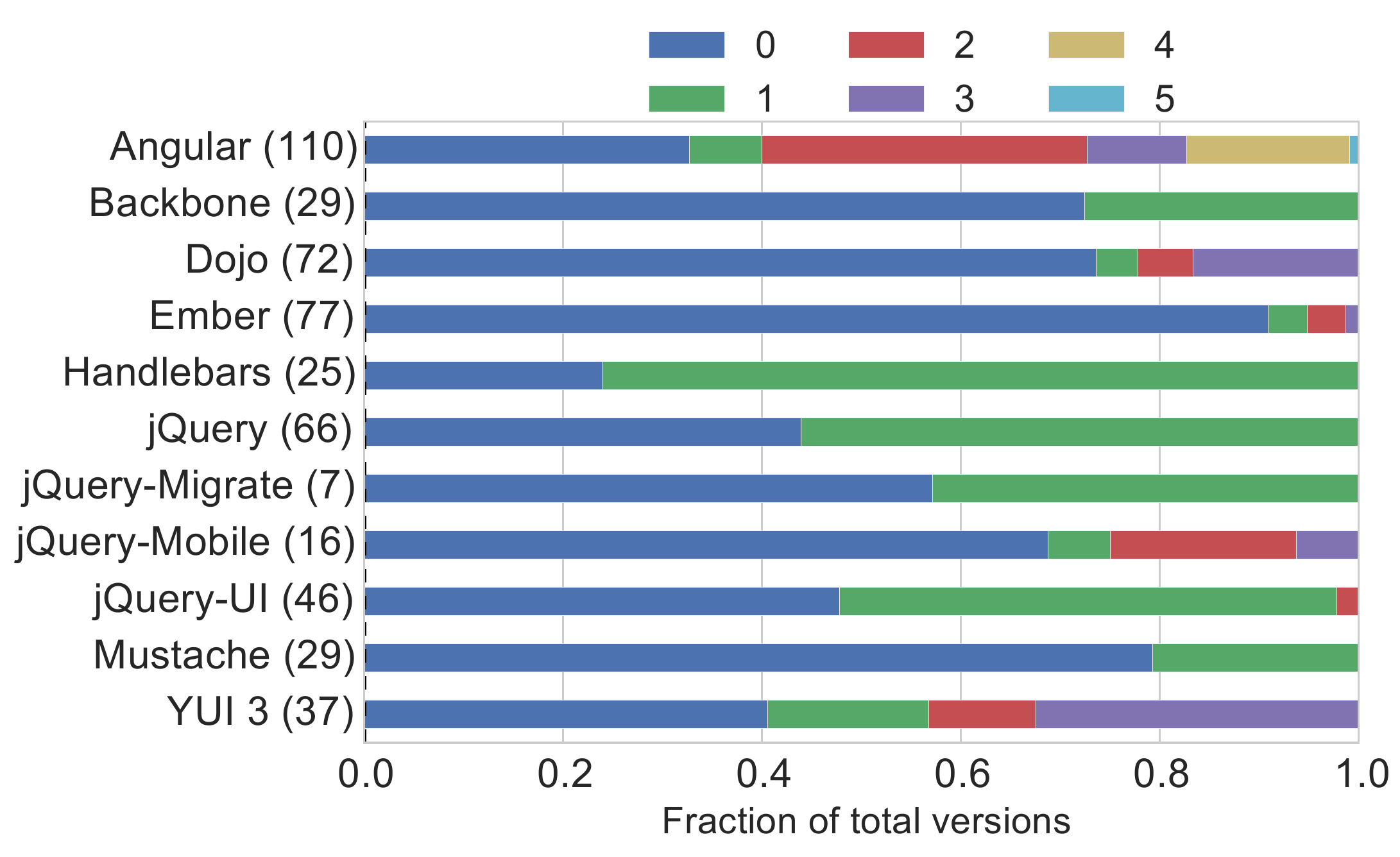}
\caption{Fraction of library versions with $i$ distinct known
         vulnerabilities each (represented by colours), out of the total
         library versions in parentheses. Angular \texttt{1.2.0} has 5
         known vulnerabilities and there are 110 versions overall.}
\label{fig:vulnsperlib}
\end{figure}

\subsubsection{Obtaining Reference Files}

Some methods of library detection require us to have access to code
samples for each version of a library. We gather library code from two
sources: the official website of each library, and from CDNs. For
the official websites, we manually download all available library versions.
However, some official websites do not provide copies of old library versions,
or they only provide copies of a subset of versions. In contrast, CDNs
typically do host comprehensive collections of old library versions in
order to not break websites that depend on older versions. We
utilise the API of one such CDN, jsDelivr, to automatically discover all available versions
of libraries on five supported CDNs.
For the remaining CDNs, we construct download link templates manually,
such as
\url{https://ajax.googleapis.com/ajax/libs/jquery/{version}/jquery.min.js}.
In doing so, we make sure that we download
all available variants of a library file,
including the full development variant and the minified production
variant without whitespace or comments.

When comparing files downloaded from official websites and different CDNs,
we noticed that even the same version and variant (\eg minified) of a library
may sometimes differ between sources. We observed
additional whitespace, removal of comments, or the
likely use of a different minifier or minifier setting, especially when
the library's developers do not provide a minified version. This observation
highlights the importance of collecting ground-truth JavaScript library
samples from as many official and semi-official sources as possible.
Therefore, we use official websites as well as
dedicated CDNs (Bootstrap
CDN and jQuery CDN), commercial CDNs (Google, Microsoft, and Yandex), and
open source CDNs (jsDelivr, cdnjs and OssCDN).

In total, we collect 81,027 JavaScript files. We analyse the sizes
of the ``main'' files of each library in our dataset (that is, we exclude
files such as plug-ins that cannot be used stand-alone), and find
that Script.aculo.us~\texttt{1.9.0} is the smallest at 996~bytes
(minified). After
accounting for duplicates and discarding files smaller than 996~bytes (to
reduce the likelihood of false positives due to shared ancillary
resources such as configuration files, localisations and plug-ins),
our final catalogue includes 19,099 distinct files.

\subsubsection{Identifying Vulnerabilities}

The last step towards building our catalogue is aggregating vulnerability
information for our 72~JavaScript libraries. Unfortunately, there is no
centralised database of vulnerabilities in JavaScript libraries;
instead, we manually compile vulnerability information from the Open
Source Vulnerability Database (OSVDB), the National Vulnerability
Database (NVD), public bug trackers, GitHub comments, blog posts, and the
vulnerabilities detected by Retire.js~\cite{retirejs}.

Overall, we are able to obtain systematically documented details of
vulnerabilities for 11 of the JavaScript libraries in our catalogue. In
some cases, the documentation for a given flaw specifies an affected
range of versions, in which case we consider all library versions within
the range to be vulnerable. In other cases, when a flaw is identified in
a specific version $v$ of a library, we consider all versions $\le v$ to
be vulnerable.

Figure~\ref{fig:vulnsperlib} shows details of the 11~libraries with
vulnerability information. For each library, we show the total number of
versions in our catalogue as well as the fraction of versions with $i$
distinct known vulnerabilities. The worst offender is Angular~\texttt{1.2.0},
which contains 5~vulnerabilities. Overall, we see that 28.3\,\%,
6.7\,\%, and 6.1\,\% of these library versions contain one, two, or
three known vulnerabilities, respectively.

\subsubsection{Limitations}
\label{sec:libcatalogue:limits}

Although we have expended a great deal of effort constructing our
catalogue of JavaScript libraries, it is impacted by several limitations.
{\em First}, by choosing GitHub for versioning and release date
information, we need to exclude a small number of libraries that have few
or no releases tagged on GitHub or do so in an apparently inconsistent
way (e.g., multiple successive releases tagged on the same day).
Furthermore, we cannot include closed-source libraries such as Google
Maps, advertising and tracking libraries like Google Analytics, and
social widgets since they typically do not publish version information.
Fortunately, the vast majority of such libraries are hosted by their
creators at a single, non-versioned URL
(\eg \url{https://www.google-analytics.com/analytics.js}), meaning that
all clients automatically include the latest version of the library.

{\em Second}, our catalogue may miss some revisions of libraries if the
author chose to patch the code and not increment the version number.
Similarly, we may miss revisions if they are denoted using non-standard
notation, such as special suffixes, four-part version numbers, etc., and
we may not possess any code samples for a version of a library if it
cannot be downloaded from the developer website or a supported CDN.

{\em Third}, our library vulnerability assessments are based solely on
publicly available documentation. We make no attempts to discover new
vulnerabilities, or to quantify the exploitability of libraries as used
on websites, for both practical and ethical reasons. Thus, although a
website may include a vulnerable library, this does not necessarily imply
that the website is exploitable. Furthermore, libraries differ in their
release cycles, attack surfaces, functionality, and public scrutiny with
respect to vulnerabilities. Thus, we do not claim to provide comparable
coverage of vulnerabilities for each library in our catalogue.

\subsection{Library Identification}
\label{sec:identification}

Identifying an unknown file as a specific version of a JavaScript library
is challenging because these libraries are text, which gives web
developers, development tools and network software the ability to modify
them, \eg by adding or removing features, concatenating multiple
libraries into a single file, or tampering with comments.
To reliably detect as many libraries as possible, we use
two complementary techniques. These techniques are conceptually similar to
those used by the {\em Library Detector} Chrome
extension~\cite{librarydetector} and Retire.js.

\subsubsection*{Static Detection} We compute the file hashes of all
observed JavaScript code and compare them to the
19,099 reference hashes in our catalogue. File hashing enables us to
identify all cases where libraries are used ``as-is.''

\subsubsection*{Dynamic Detection} During the crawl, we detect the
presence of libraries in the browser by fingerprinting the JavaScript
runtime environment and by relying on libraries to identify
themselves. Specifically, modern libraries typically make themselves
available to the environment by means of a global variable that can
be detected at runtime. Furthermore,
most libraries in our catalogue contain a variable or method that returns
the version of the library. As an illustration, the following snippet of
JavaScript code detects jQuery:

\begin{lstlisting}
var jq = window.jQuery || window.$ || window.$jq || window.$j;
if(jq && jq.fn) {
   return jq.fn.jquery || null; //version (if known)
} else {
   return false; //jQuery not found
}
\end{lstlisting}
\vspace{-2em}
Line~1 extracts jQuery's global variable, and line~3 returns the version
number if it exists in its \texttt{fn.jquery} attribute. Note that in
order to prevent false positives, we check for the global variable
\textit{and} that the \texttt{fn} attribute exists.
Later on, we discard all detections with missing or syntactically
invalid version strings.

While this dynamic methodology detects libraries even if the source code
has been (lightly) modified, it relies on the version attribute to be
present. Hence, we can dynamically detect only 39 out of the 72
libraries, and for some, we do not detect (typically older) versions
lacking the version string.
Table~\ref{tab:libraries} compares the versions detected dynamically in
our crawls \textcircled{d} to our static reference catalogue
\textcircled{s}.
Version coverage is often similar; dynamic outperforms static when
CDNs are incomplete.

\subsubsection*{Limitations} Our two detection techniques represent a
best-effort approach to identifying JavaScript libraries in the wild.
However, there are cases where both techniques can fail. For example,
heavily modified libraries will not match our file hashes nor will they
match the dynamic signatures. Furthermore, we rely on the correctness of
our information sources, \ie that CDNs contain the version of a
library that they claim, and that libraries export the
correct version string and do not attempt to conceal their presence.
Effectively, these limitations mean that our measurement results should
be viewed as lower bounds.

\subsection{Data Collection}
\label{sec:collection}
A central contribution of our work is to analyse not only whether outdated
libraries are being used, but \textit{why} this may be the case. This
implies that detecting whether a library exists in a window or frame is
not enough; we must also detect if it was loaded by another script.
To model
causal inclusion relationships of resources in websites, we introduce the
theoretical concept of \textit{causality trees} and implement it in a
modern browser.
We integrate our two library detection methods into this modified browser
environment and use it to collect
data about the usage of JavaScript libraries
on the Web.

\subsubsection*{Causality Trees} The goal of a causality tree is to
represent the causal element creation relationships that occur during the
loading and execution of a dynamic website in a modern browser. A
causality tree contains a directed edge $A \rightarrow B$ if and only if
element $A$ causes element $B$ to load. More specifically, the elements
we model include scripts, images and other media content, stylesheets, and
embedded HTML documents. A relationship exists whenever an element
creates another element (\eg a script creates an iframe) or
changes an existing element's URL (\eg a script changes the URL of an
iframe or redirects the main document), which is equivalent to
creating a new element with a different URL.

While the nodes in a causality tree correspond to nodes in the website's
DOM, their structure is entirely unrelated to the hierarchical DOM tree.
Rather, nodes in the causality tree are snapshots of elements in the DOM
tree {\em at specific points in time}, and may appear multiple times if
the DOM elements are repeatedly modified. For instance, if a script creates an
iframe with URL $U_1$ and later changes the URL to $U_2$, the
corresponding script node in the causality tree will have two document
nodes as its children, corresponding to URLs $U_1$ and $U_2$,
but referring to the same HTML \texttt{<iframe>} element.
Similarly, the predecessor of a node in the causality tree is not
necessarily a predecessor of the corresponding HTML element in the DOM
tree; they may even be located in two different HTML documents, such as
when a script appends an element to a document in a different frame.

\begin{figure}[t]
\centering
\includegraphics[width=1.0\columnwidth]{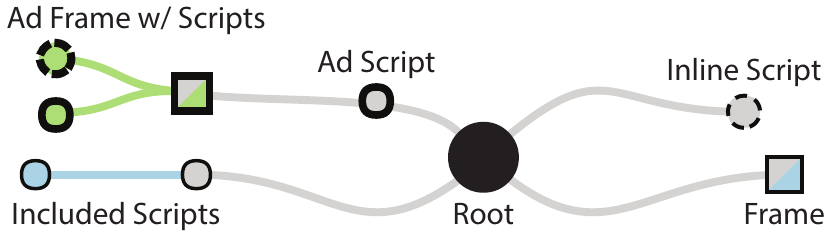}
\caption{Example causality tree.}
\label{fig:examplecaus}
\end{figure}

Figure~\ref{fig:examplecaus} shows a synthetic example of a causality
tree. The large black circle is the document root (main document), filled
circles are scripts, squares are HTML documents (\eg embedded in frames
or corresponding to a new main document if there is a top-level redirect),
and empty circles are other resources (\eg images). Edges denote
``created by'' relationships; for example, in Figure~\ref{fig:examplecaus}
the main document included the grey script, which in turn included the
blue script. Dashed lines around nodes denote inline scripts, while solid lines
denote scripts included from an URL. Thick outlines denote that a
resource was included from a known ad network, tracker, or social widget
(see below for more details).

The colour of nodes in Figure~\ref{fig:examplecaus} denotes which
document they are {\em attached to} in the DOM: grey corresponds to
resources attached to the main document, while we assign one of four
colours to each further document in frames. Document squares contain the
colour of their parent location in the DOM, and their own assigned colour.
Resources created by a script in one frame can be attached to a document
in another frame, as shown by the grey script which has a blue child in
Figure~\ref{fig:examplecaus}, \ie the blue script is a child of the blue
document in the DOM.

Figure~\ref{fig:mercantile} shows the causality tree of
\texttt{mercantil.com}, with images and other irrelevant node types
omitted for clarity. It includes three clearly visible social media
widgets: Twitter, Facebook, and LinkedIn.
Note that the web developer embedded code provided by the social
networks into the main document, which in turn initialises each widget
and creates one or more frames for their contents. We also see that the
causality tree includes multiple copies of jQuery in the main document,
which we will discuss in detail in Section~\ref{sec:analysis:duplicates}.

\subsubsection*{Implementation of Causality Trees}
Our definition of causality trees is related to the
concepts previously used to analyse malicious JavaScript
inclusions~\cite{arshad-2016-fc} and cookie matching between online ad
exchanges~\cite{bashir-2016-usenix}, but it differs in the details. Our
implementation is independent from the aforementioned works and uses a
different technological approach by building on the Chrome Debugging
Protocol~\cite{chromeDebugging}
to minimise the necessity for brittle browser source code modifications.

The Chrome Debugging Protocol provides programmatic access to the browser
and allows clients to attach to open windows, inspect network
traffic, and interact with the JavaScript environment and the DOM
tree loaded in the window. Two prominent uses of this API are the Chrome
Developer Tools (an HTML and JavaScript front-end to the
protocol) and Selenium's WebDriver interface to remotely control Chrome.

At a high level, we generate causality trees by observing resource
requests through the network view of the debugging protocol. Note that
this view includes resources not actually loaded over the network, \eg
inline URL schemas such as \texttt{data:} or \texttt{javascript:}. We
disable all
forms of caching to observe even duplicate resource inclusions within the
same frame, which are otherwise handled through an in-memory cache. For
each loaded resource, the protocol allows us to identify the frame in
which the resource is located as well as the initiating script, where
applicable. Similarly, we utilise protocol methods to be notified of
script generation events and store the source code of both inline and
URL-based JavaScript. (This includes source code from attribute-based
event handlers and string evaluation, which we both model as inline
script nodes.)
We store a log of all relevant events during the crawl and assemble the
causality trees in a post-processing step.

\subsubsection*{Integration of Library Detection}
Hash-based detection of libraries is relatively straight-forward to
integrate with our crawler; we simply compute the source code hashes of
all script nodes in the causality trees and look them up in our reference
catalogue during post-processing.

Integrating dynamic library detection is more challenging---out of the
box, existing detection methods can only detect whether a JavaScript
context (window or frame) contains a library or not, but in general this
information is not sufficient to properly label the correct script node
in the causality tree. The Chrome Debugging Protocol allows us to link a
method executed in a JavaScript context to the script that contains its
implementation; thus, once we hold a reference to a JavaScript library
object (such as \texttt{jq} in the example from
Section~\ref{sec:identification}), we dynamically enumerate its instance
methods
and use an arbitrary one of them to identify the implementing script.

Another challenge in dynamically detecting libraries during the crawl is
that we need to inject the detection code into each frame of a website,
since each frame has its own JavaScript scope and may contain
independent library instances. This is further complicated by the fact
that modern websites are quite dynamic and an ad frame, for instance, may
quickly navigate to a different URL, which causes any library previously
loaded in the frame to be unloaded or replaced. Lastly, we observe
that many websites
include multiple copies of the same library, including different versions
of the same library (refer to our analysis in
Section~\ref{sec:analysis:duplicates} for details). Typically, only one library
instance can exist in each context because the more recently loaded
instance replaces the previous global reference. 

In order to be able to study these phenomena in detail and
also address the other aforementioned detection challenges, we
inject the detection code into each frame and execute it
every 4~seconds. Note that it would not be feasible from a performance
point of view to execute the detection code after each script creation
event since websites routinely contain thousands of script nodes (see
Section~\ref{sec:analysis:general}).

To detect cases of inclusions that we may miss due to the
four-second detection interval, we additionally execute the dynamic
detection code {\em post hoc} on all scripts found during the crawl.
This execution is done
in an individual Node.js environment with a fake DOM tree. This offline
detection step cannot fully replace in-browser detections, however, since
some libraries such as jQuery UI have code or environment prerequisites
that cause the offline detection to fail.

\subsubsection*{Annotation of Ads, Trackers and Widgets}
To further clarify the provenance of scripts we observe in our crawl,
we aim to determine whether they are related to known advertising,
tracking, or social widget code. We achieve this by injecting a customised
version of the AdBlock browser extension into each frame. Our version of
AdBlock flags content but permits it to load, and we verified that
our customised version remains undetected by common ad-block detection
scripts. We use EasyList and EasyPrivacy to identify advertising and
tracking content, and Fanboy's Social Blocking List to detect social
network widgets. For our analysis, we label an element in the causality
tree as ad/tracker/widget-related whenever the corresponding element or
any parent in the DOM tree is labeled by AdBlock. Additionally, we
propagate these labels downwards to all children of the labelled node
in the causality tree.

\subsubsection*{Crawl Parameters}
To gain a representative view of JavaScript library usage on the Web,
we collected two different datasets. {\em First}, we crawled the Alexa
Top~75\,k domains, which represent websites popular with users.
{\em Second}, we crawled 75\,k domains randomly sampled from a snapshot
of the \texttt{.com} zone, that is, a random sample of all websites with
a \texttt{.com} address, which we expect to be dominated by less popular
websites. We conducted the two crawls in May~2016 from IP addresses in
a /24~range in the US. We observe $\sim$5\,\% and
$\sim$17.2\,\%~failure rates in \textsc{Alexa} and \textsc{Com}, leaving
us with data from 71,217 and 62,086 unique domains, respectively.
Failures were due to timeouts and unresolvable domains, which is expected
especially for \textsc{Com} since the zone file contains domains that may
not have an active website.

To preserve the fidelity of our data collection, our crawler is based
on Chromium and includes support for Flash. We disable various security
mechanisms such as malware and phishing filters.
We only crawl the homepage of each visited site due to the presence of many
sites that thwart deeper traversal by requiring log-ins. While visiting a
page, the crawler scrolls downwards to trigger loading of any dynamic
content. As we found page-loaded events to be unreliable, our crawler
remains on each page for a fixed delay of 60~seconds before clearing its
entire state, restarting, and then proceeding to the next site.

\subsection{Validation}

As the final step in our methodology, we validate that our static and
dynamic detection methods work in practice.

\subsubsection*{Dynamic Detection}
To investigate the efficacy of our dynamic detection code, we conduct
a controlled experiment: we load each of the reference libraries in
our catalogue into Node.js, one at a time, and attempt to detect each
file with the dynamic detection method. Intuitively, we know the exact
version of each loaded library, which enables us to assess the accuracy
of the dynamic detection.

Overall, we observe that the dynamic detection code is
able to identify the exact name and version of 79.2\,\% of the
libraries, as well as the name (but not the version) of 18.6\,\% of the
libraries. Only 2\,\% of libraries fail to be identified (\ie were
false negatives). We manually examine the libraries that are only
detected by name, and find that the vast majority are older versions
that do not include a variable or method that returns the library
version. Of course, 100\,\% of these libraries can be detected based
on their file hashes, which reinforces the importance of using multiple
techniques to identify libraries.

\subsubsection*{Static Detection}
To investigate the efficacy of library detection using file hashes,
we conduct a second controlled experiment: we randomly select from
our crawled data 415 unique scripts that the dynamic detection code
classifies as being jQuery, and attempt to detect them based on
their file hashes. In this case, we are treating the output of the
dynamic detection method as ground truth.

Overall, we observe that only 15.4\,\% of the libraries can be
identified as jQuery based on their file hash. Although this is a low
detection rate, the result also matches our expectation that developers
often deploy customised versions of libraries. For example, 90\,\% of
the jQuery libraries that we fail to detect via file hashing
contain fewer than 150 line break characters, whereas non-minified copies
of jQuery from our catalogue contain more than 1900. This strongly suggests
that the unique scripts are custom-minified versions of jQuery.

\subsubsection*{Hypothetical ``Name-in-URL'' Detection}
For the last validation step, we consider a simple library detection
heuristic. The heuristic flags a script file as jQuery, for instance,
whenever the string ``jquery'' appears in the URL of the script.
To evaluate the accuracy of this heuristic, we extract from our
\textsc{Alexa} crawl the set of script URLs that contain ``jquery,''
and the URLs of scripts detected as jQuery by our dynamic and static
methods. Out of these URLs, 22.3\,\% contain ``jquery'' and are also
detected as the library; 69\,\% are flagged only by the heuristic, and
8.8\,\% are detected only by our dynamic and static methods.
The heuristic appears to cause a large number of false positives due
to scripts named ``jquery'' without containing the library, and it also
seems to suffer from false negatives due to scripts that contain jQuery
but have an unrelated name.

We validate this finding by manually examining 50 scripts from each of
the two set differences.
The scripts in the detection-only sample appear to contain additional
code such as application code or other libraries. Only one of these
scripts does not contain jQuery but Zepto.js, an alternative to jQuery
that is partially compatible and also defines the characteristic
\texttt{\$.fn} variable.
On the other hand, none of the scripts in the heuristic-only sample
can be confirmed as the library; nearly all of them contain
\textit{plug-ins} for jQuery but not the library itself.

The results for the Modernizr library, which does not have an equivalent to
jQuery's extensive plug-in ecosystem, confirm this trend. The overlap
between heuristic and detection is 55.3\,\% of URLs, heuristic-only
0.8\,\%, and detection-only 44\,\%---%
the simple heuristic misses many Modernizr files renamed by developers.
These results underline the need for more robust detection techniques
such as our dynamic and static methods; we do not use the heuristic in
our analysis.

\begin{figure*}[!htbp]
  \begin{minipage}[b]{0.322\textwidth}
    \capstart
    \includegraphics[width=1\columnwidth]{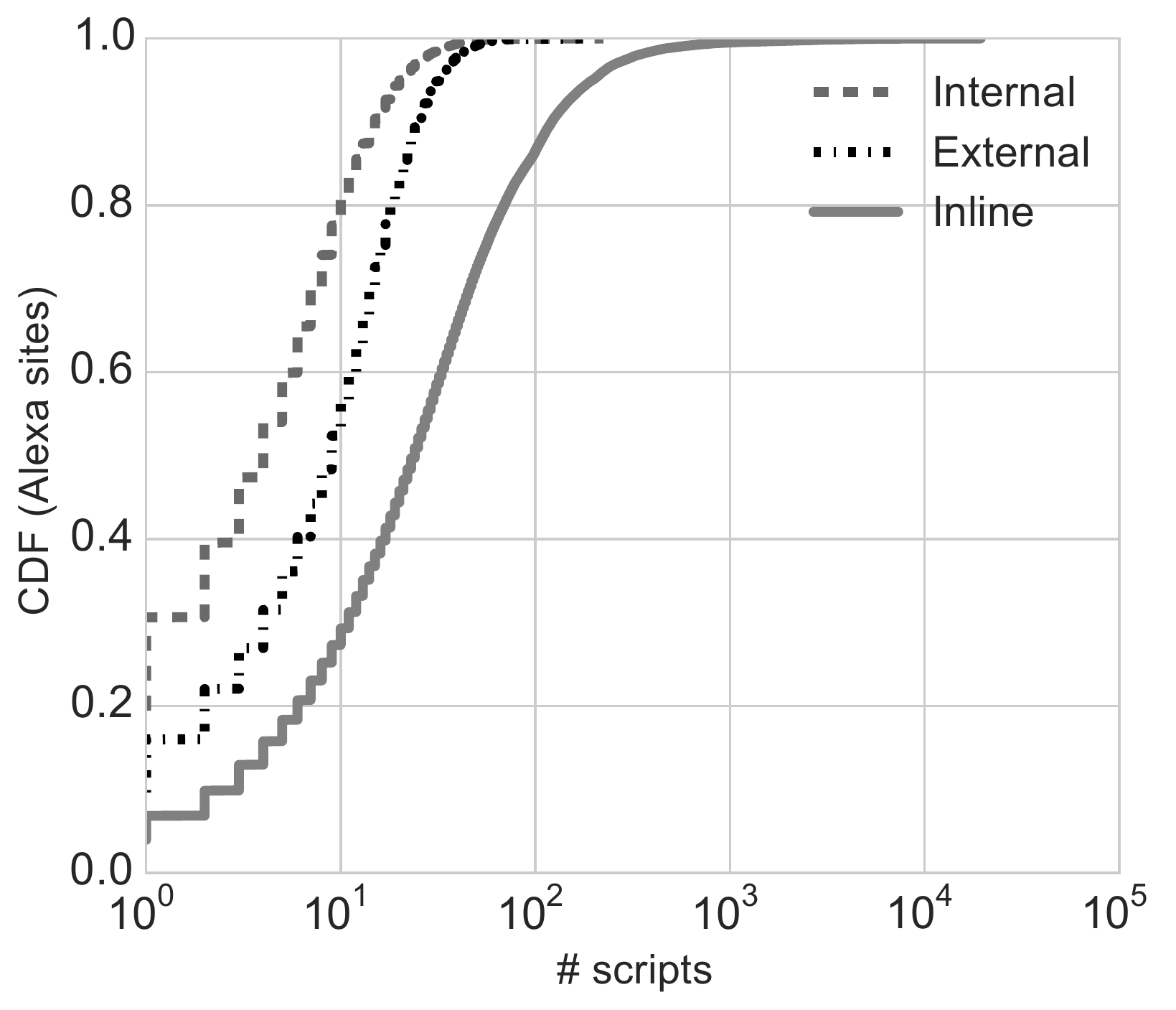}
    \caption{Distribution of JavaScript inclusion type frequency per site
             in \textsc{Alexa}.}
    \label{fig:distrib_all_js_alexa}
  \end{minipage}
\hfill
  \begin{minipage}[b]{0.33\textwidth}
    \capstart
    \includegraphics[width=1\columnwidth]{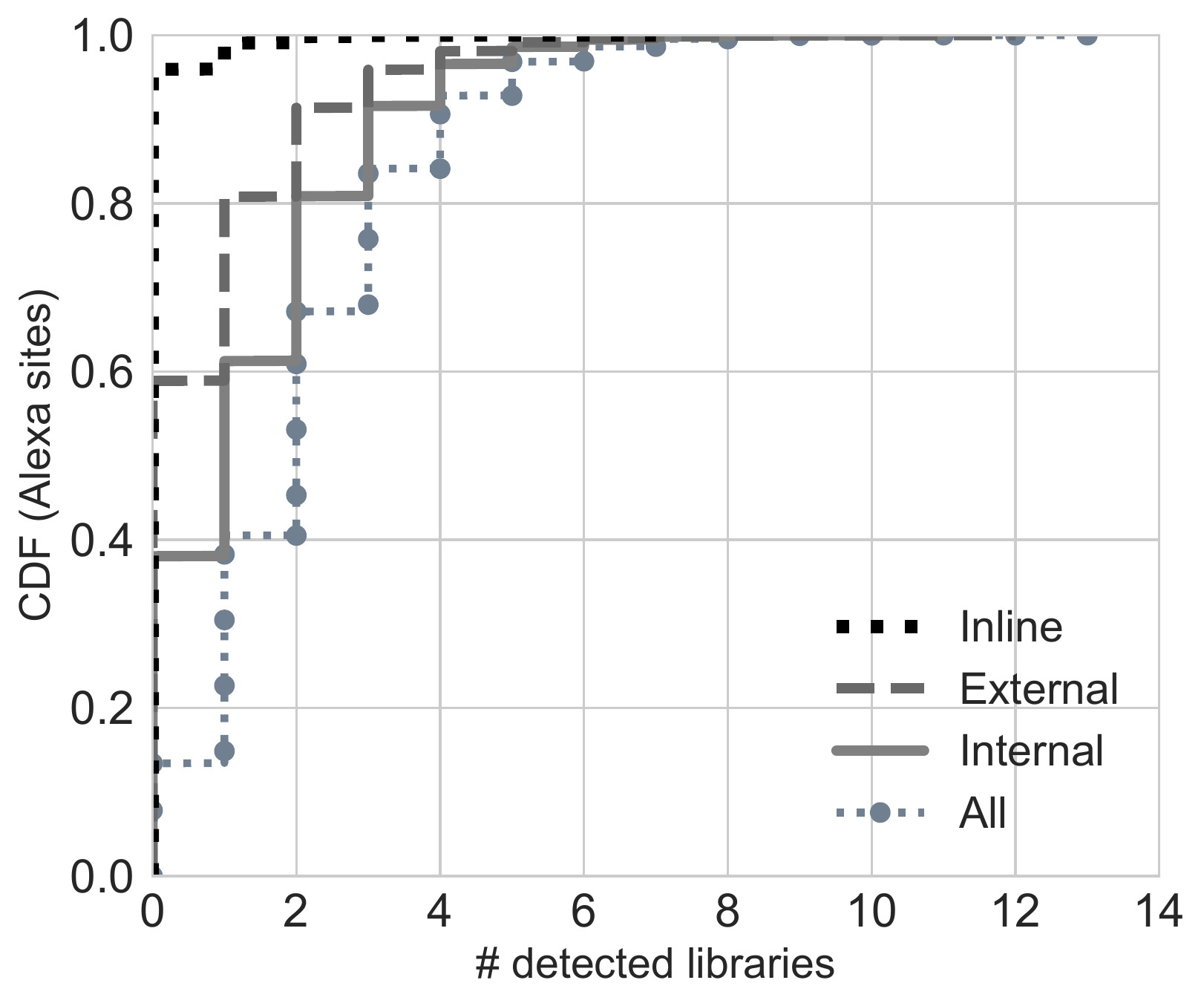}
    \caption{Distribution of library inclusion type frequency per site
             in \textsc{Alexa}.}
    \label{fig:ditrib_detected_alexa}
  \end{minipage}
\hfill
  \begin{minipage}[b]{0.32\textwidth}
    \centering
    \capstart
    \includegraphics[width=1.0\columnwidth]{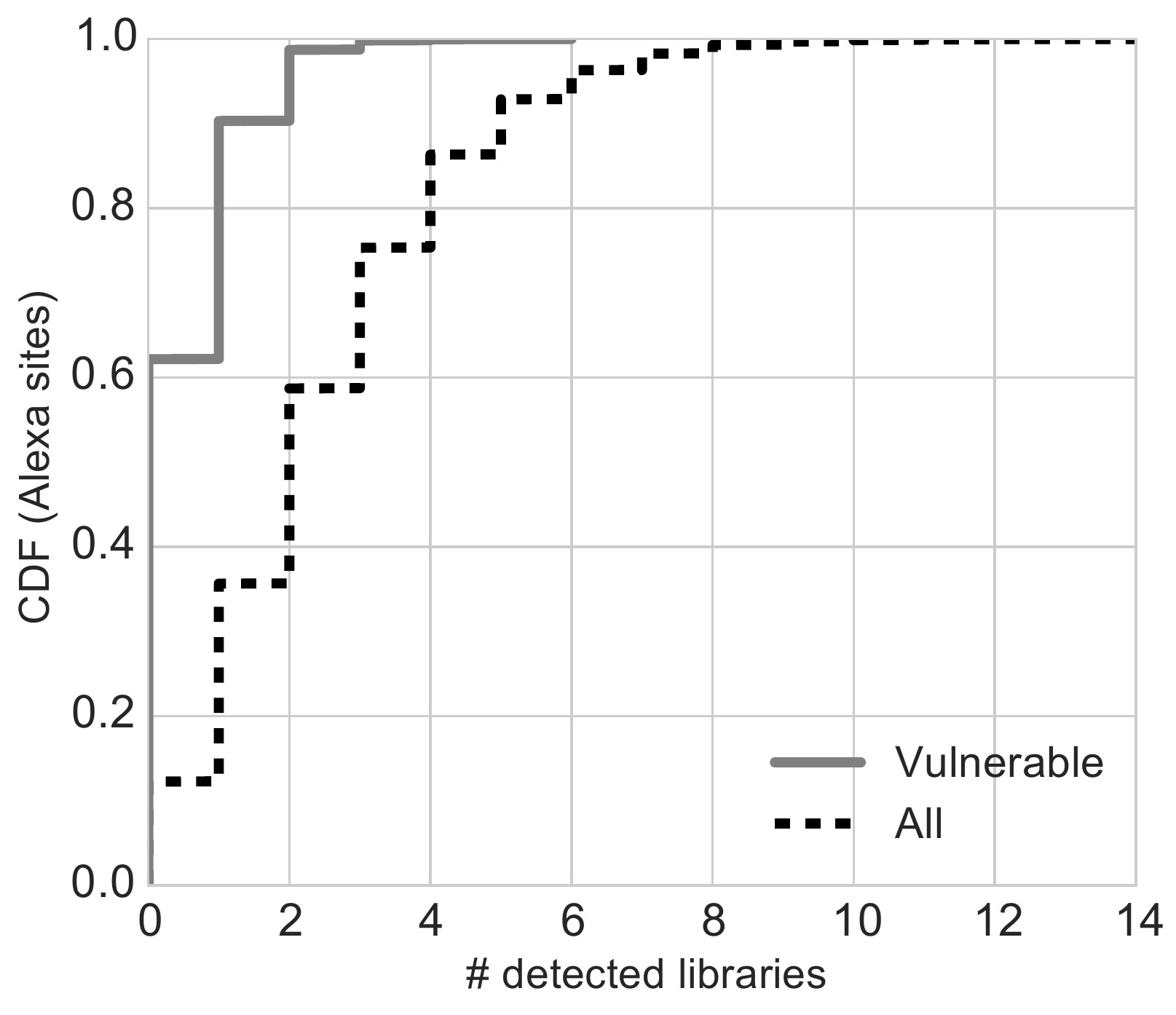}
    \caption{Distribution of vulnerable library count versus overall
             library count per site in \textsc{Alexa}.}
    \label{fig:distrib_vuln_alexa}
  \end{minipage}
\end{figure*}

\begin{figure*}[!htbp]
  \begin{minipage}[b]{0.322\textwidth}
    \capstart
    \includegraphics[width=1\columnwidth]{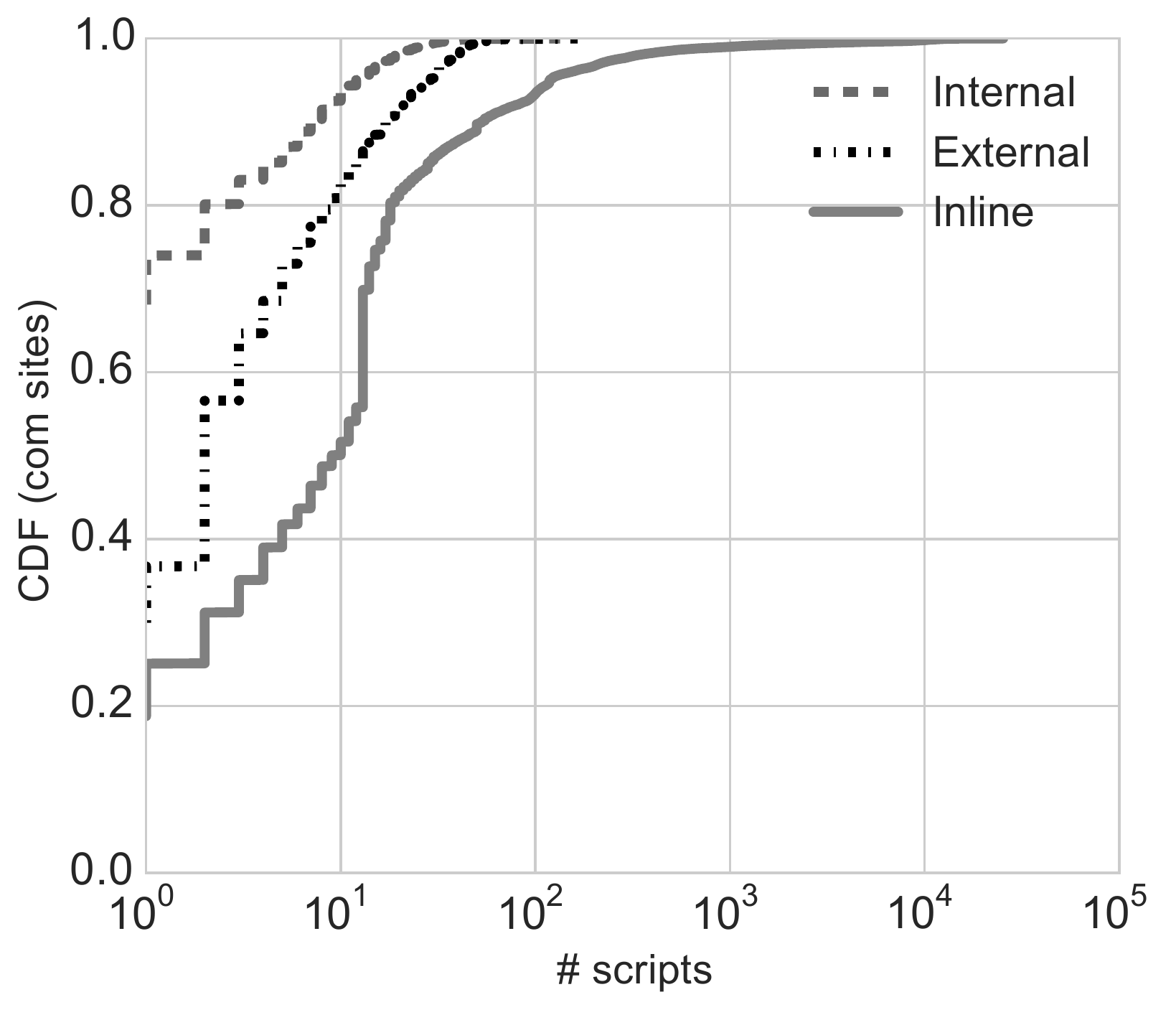}
    \caption{Distribution of JavaScript inclusion type frequency per site
             in \textsc{Com}.}
    \label{fig:distrib_all_js_com}
  \end{minipage}
\hfill
  \begin{minipage}[b]{0.33\textwidth}
    \capstart
    \includegraphics[width=1\columnwidth]{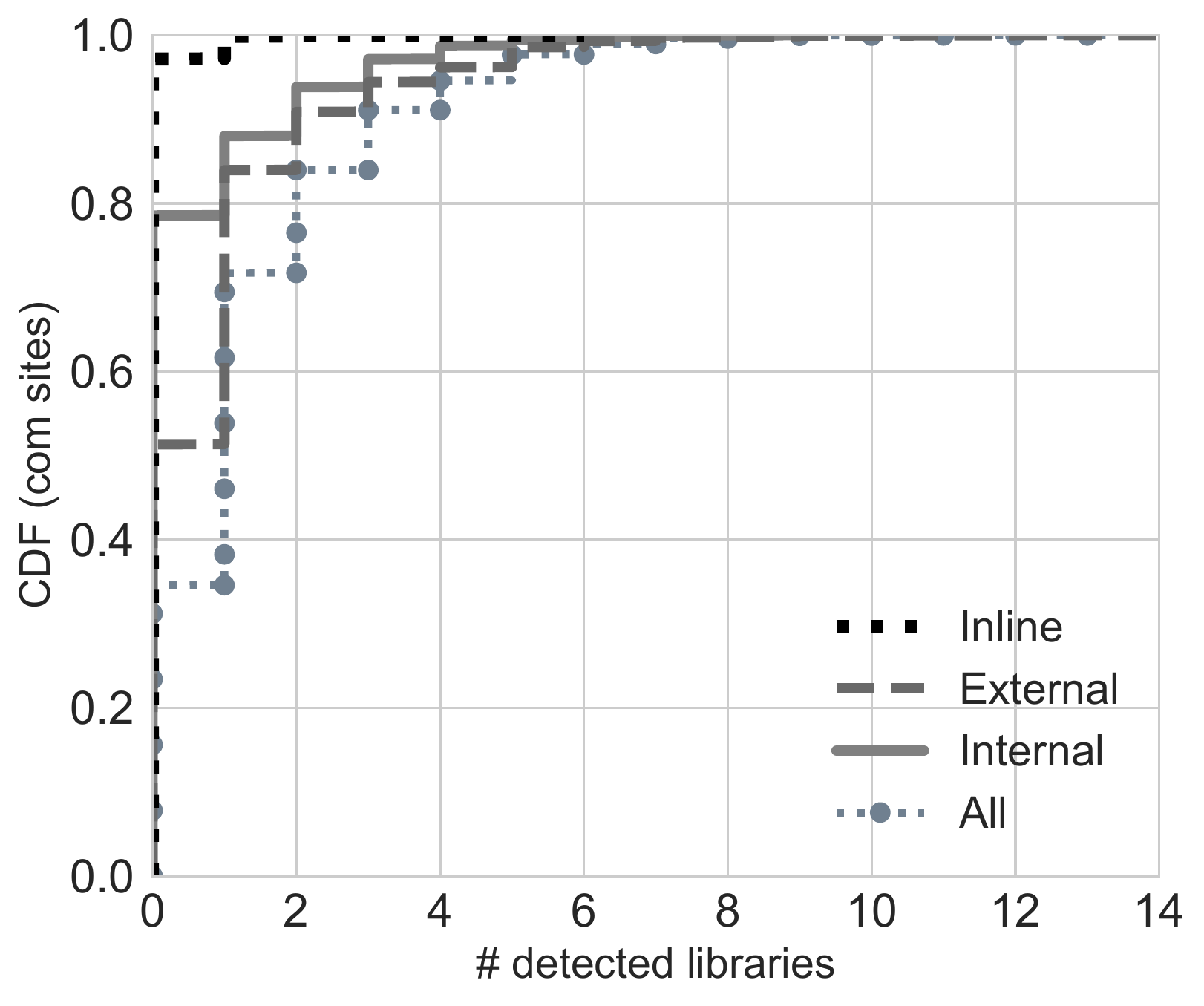}
    \caption{Distribution of library inclusion type frequency per site
             in \textsc{Com}.}
    \label{fig:ditrib_detected_com}
  \end{minipage}
\hfill
  \begin{minipage}[b]{0.32\textwidth}
    \centering
    \capstart
    \includegraphics[width=1.0\columnwidth]{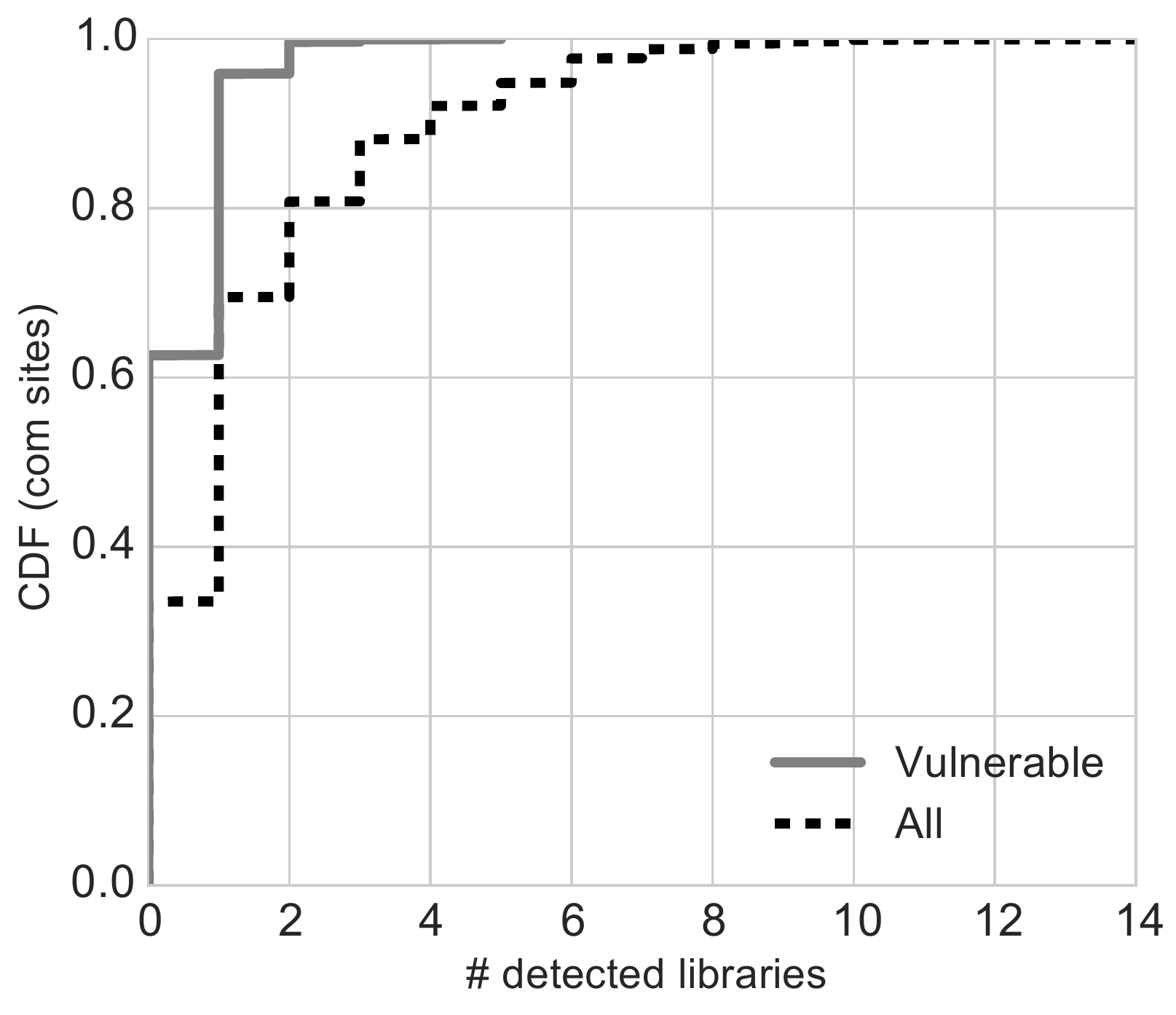}
    \caption{Distribution of vulnerable library count versus overall
             library count per site in \textsc{Com}.}
    \label{fig:distrib_vuln_com}
  \end{minipage}
\end{figure*}

\section{Analysis}
\label{sec:analysis}

In this section, we analyse the data from our web crawls. First, we present a general overview
of the dataset by drilling down into the causality trees, overall JavaScript inclusion statistics,
and vulnerable JavaScript library inclusions. Next, we examine risk factors for sites that include
vulnerable libraries, and the age of vulnerable libraries (\ie relative to the latest release of
each library). Finally, we examine unexpected, duplicate library inclusions in websites, and 
investigate whether common remediations practices are useful and used in practice.

\subsection{Causality Trees}
\label{sec:analysis:causalityTrees}

We begin our analysis by measuring the complexity of the
websites in our crawls. The median causality tree in \textsc{Alexa}
contains 133~nodes (p95:~425, max:~19,508) whereas it is 38~nodes
(p95:~298, max:~25,485) in \textsc{Com}, indicating that
\textsc{Alexa} contains a larger share of more complex websites.
Similarly, \textsc{Alexa} and \textsc{Com} contain a median of 52 and
14~image nodes, 44 and 16~script nodes, and 4 and 2~document nodes per
causality tree, respectively.
95\,\% of frames in both crawls show only one
or two documents during the 60\,s of our crawl, whereas a few cycle
through larger numbers of documents (max:~12 for \textsc{Alexa} and 32
for \textsc{Com}).

The median depth of the causality trees (defined as the length of the
longest path from the root to a leaf) is 4 inclusions (p95:~9, max:~438)
in \textsc{Alexa} and 3 (p95:~8, max:~62) in \textsc{Com}. Since paths
correspond to causal relationships, intuitively this means that the
inclusion of a node could have been influenced by up to 438
predecessors. In both crawls, images tend to appear further up in the
causality trees at a median depth of 1, that is, at least half of them
are directly included in the main document, whereas documents tend to
appear further down at a median depth of 2, which indicates that they
are more frequently dynamically generated.

\subsection{General JavaScript Statistics}
\label{sec:analysis:general}

\begin{table}[t]
\centering
\scriptsize

\begin{minipage}{0.48\textwidth}
\caption{External JavaScript: Top~10 market share. \hspace{0.2cm} Sorted by
         \textsc{Alexa}; data omitted for hosts not part of the Top~10.}
\centering
  \begin{tabular}{l@{\hspace{0.15cm}}rr}
    \toprule
	\textbf{Hostname} & \textbf{\textsc{Alexa}} & \textbf{\textsc{Com}} \\
    \midrule
    \texttt{www.google.com} &  & $11.5$\,\% \\
	\texttt{www.google-analytics.com} & $9.5$\,\% & $9.4$\,\% \\
	\texttt{ak2.imgaft.com} &  & $5.6$\,\% \\
	\texttt{ajax.googleapis.com} & $3.1$\,\% & $3.9$\,\% \\
	\texttt{pagead2.googlesyndication.com} & $2.9$\,\% & $1.5$\,\% \\
	\texttt{connect.facebook.net} & $2.8$\,\% & $1.5$\,\% \\
	\texttt{www.googletagmanager.com} & $2.4$\,\% &  \\
	\texttt{www.googletagservices.com} & $2.1$\,\% &  \\
	\texttt{partner.googleadservices.com} & $2.0$\,\% &  \\
	\texttt{www.googleadservices.com} & $2.0$\,\% &  \\
	\texttt{platform.twitter.com} & $1.8$\,\% & $1.2$\,\% \\
	\texttt{apis.google.com} & $1.5$\,\% & $1.3$\,\% \\
	\texttt{maps.googleapis.com} &  & $1.3$\,\% \\
	\texttt{s.ytimg.com} &  & $1.2$\,\% \\
    \bottomrule
  \end{tabular}
\label{tab:hostRankingJavascript}
\vspace{0.77cm}
\end{minipage}

\begin{minipage}{0.48\textwidth}
\caption{Detected JS libraries: Top~10 market share. Sorted by
         \textsc{Alexa}; data omitted for hosts not part of the Top~10.}
\centering
\scriptsize
  \begin{tabular}{l@{\hspace{1cm}}rr}
    \toprule
	\textbf{Hostname} & \textbf{\textsc{Alexa}} & \textbf{\textsc{Com}} \\
    \midrule
	\texttt{ak2.imgaft.com} &  & $18.3$\,\% \\
	\texttt{ajax.googleapis.com} & $17.6$\,\% & $12.7$\,\% \\
	\texttt{code.jquery.com} & $3.5$\,\% & $3.1$\,\% \\
	\texttt{static.hugedomains.com} &  & $2.8$\,\% \\
	\texttt{static.parastorage.com} &  & $2.3$\,\% \\
	\texttt{img.sedoparking.com} &  & $1.6$\,\% \\
	\texttt{img4.wsimg.com} &  & $1.5$\,\% \\
	\texttt{cdnjs.cloudflare.com} & $2.1$\,\% & $1.4$\,\% \\
	\texttt{static.squarespace.com} &  & $1.3$\,\% \\
	\texttt{s0.2mdn.net} & $1.4$\,\% & \\
	\texttt{ivid-cdn.adhigh.net} & $1.2$\,\% & \\
	\texttt{maxcdn.bootstrapcdn.com} & $0.9$\,\% & \\
	\texttt{s0.wp.com} &  & $0.9$\,\% \\
	\texttt{cdn10.trafficjunky.net} & $0.4$\,\% & \\
	\texttt{netdna.bootstrapcdn.com} & $0.4$\,\% & \\
	\texttt{ajax.aspnetcdn.com} & $0.3$\,\% & \\
	\texttt{dsms0mj1bbhn4.cloudfront.net} & $0.3$\,\% & \\
    \bottomrule
  \end{tabular}
\label{tab:hostRankingLibraries}
\end{minipage}

\end{table}

Scripts are the most common node type in our causality
trees; 97\,\% of \textsc{Alexa} sites and 83.6\,\% of \textsc{Com}
sites contain JavaScript. The most common script type are
\textit{inline} scripts, which includes script code embedded
as text in a \texttt{<script>} tag, any code in
attribute-based event handlers (such as the \texttt{onclick} attribute),
and any code evaluated from strings using methods such as
\texttt{eval()}. The causality trees from the
\textsc{Alexa} and \textsc{Com} crawls contain a median of 24 and 9
inline scripts, respectively, with 13.6\,\% and 6.7\,\% of the sites having hundreds
of inline scripts---the maximum observed was 19\,K and 25\,K.

When looking at URL-based script inclusions, we distinguish between
\textit{internal} scripts, that is, code hosted on the same domain as
the website (or a subdomain of the website), and all other scripts that we call \textit{external}.
Figures~\ref{fig:distrib_all_js_alexa} and \ref{fig:distrib_all_js_com}
depict the distribution of script inclusion types for \textsc{Alexa}
and \textsc{Com}, respectively.
About 91.7\,\% of all \textsc{Alexa} sites include at least one
external script, which is consistent with the 88.5\,\% Nikiforakis et
al.\ found for the Alexa Top~10\,k~\cite{nikiforakis-2012-ccs}.
The median number of internal scripts is 4 per site in \textsc{Alexa}
(0~in \textsc{Com}) and the maximum is 224 (134). Externally-hosted
script appears more prevalent with a median of 9 in \textsc{Alexa}
(max: 202) and 2 in \textsc{Com} (max: 172); 6.3\,\% of sites in
\textsc{Alexa}, and 4.8\,\% in \textsc{Com}, include 30 or more external
scripts. However, we note
that we likely underestimate the fraction of internal scripts because
our domain name-based heuristic cannot infer that two domains may be
under the same administrative control when one is not a subdomain of
the other. In both crawls, the median inclusion depth of an internal
script node is 1, \ie directly included by the main document,
whereas the median depth of an external script node is 2, suggesting
indirect inclusions.

Each site includes scripts from a median of 5
external hosts in \textsc{Alexa} (2 in \textsc{Com}) with a maximum
of 50 (45). Table~\ref{tab:hostRankingJavascript} lists
the hostnames most frequently included by \texttt{<script>} tags
(counting each hostname at most once per site to reduce bias through
multiple inclusions). At least 5 of the Top~10
in \textsc{Alexa} are related to advertising, especially Google's ad
platforms. This agrees with prior work that has found Google's Web
advertising presence to be nearly ubiquitous
(\cite{gill-imc13,cahn-www16}).

\subsection{General Library Statistics}
\label{sec:analysis:detections}
Next, we narrow our analysis to focus just on JavaScript libraries from
our catalogue. We use the union of detections made by all our techniques,
but exclude detections where the version of the library is unknown.
Overall, we detect at least one of our 72 target
libraries on 86.6\,\% of all \textsc{Alexa} sites and 65.4\,\% of
all sites in \textsc{Com}. Table~\ref{tab:libraries} lists the 30
libraries that appear on most sites in \textsc{Alexa}. We also list the
percentage of sites in \textsc{Com} that include the same libraries;
however, for clarity we omit two libraries (both below 0.5\,\%) that are
part of the Top~30 in \textsc{Com} but not in \textsc{Alexa}.
jQuery is by far the most popular library,
which we find on 83.9\,\% of the \textsc{Alexa} sites and 61.1\,\%
of the \textsc{Com} sites. SWFObject, a library used to include Adobe
Flash content, is ranked seventh (3.7\,\%) and tenth (2.4\,\%)
despite
being discontinued~\cite{swfobject}
since 2013. On the other hand, several relatively well-known libraries
such as D3, Dojo and Leaflet appear below the Top~30 in both crawls.
We also observe that the market share for these libraries is generally
consistent with the data reported by Wappalyzer.

\begin{table}[t]
\caption{Inclusion types of detected libraries: loaded from
inline/internal/external script \& inclusions (of any type)
due to ad/tracker/widget code. Greyed out if <100 inclusions;
omitted if <10. Libraries counted at most once
per site.}
\vspace{0.07cm}
\centering 
\scriptsize

\begin{tabular}{l@{\hspace{0.2cm}}rrrrrrrrr}
\toprule
 & \multicolumn{4}{c}{\textbf{\textsc{Alexa}}} & \multicolumn{4}{c}{\textbf{\textsc{Com}}}\\
\cmidrule(lr){2-5} \cmidrule(lr){6-9}
 & \multicolumn{3}{c}{\textbf{Included From}} & \textbf{Ad/Tr./} & \multicolumn{3}{c}{\textbf{Included From}} & \textbf{Ad/Tr./} \\
\textbf{Library} & \textbf{Inl.} & \textbf{Int.} & \textbf{Ext.} & \textbf{Widget} & \textbf{Inl.} & \textbf{Int.} & \textbf{Ext.} & \textbf{Widget} \\
\midrule
jQuery & \textcolor{black}{2.7} & \textbf{\textcolor{black}{58.8}} & \textcolor{black}{38.6} & \textcolor{black}{3.4} & \textcolor{black}{2.9} & \textcolor{black}{29.5} & \textbf{\textcolor{black}{67.6}} & \textcolor{black}{25.6} \\
Modernizr & \textcolor{black}{3.0} & \textbf{\textcolor{black}{65.4}} & \textcolor{black}{31.6} & \textcolor{black}{12.2} & \textcolor{gray}{1.4} & \textcolor{black}{38.2} & \textbf{\textcolor{black}{60.4}} & \textcolor{black}{3.5} \\
Bootstrap & \textcolor{black}{1.7} & \textbf{\textcolor{black}{69.9}} & \textcolor{black}{28.4} & \textcolor{gray}{0.3} & \textcolor{gray}{0.5} & \textcolor{black}{49.2} & \textbf{\textcolor{black}{50.3}} & \textcolor{gray}{3.5} \\
SWFObject & \textcolor{black}{4.4} & \textcolor{black}{34.2} & \textbf{\textcolor{black}{61.4}} & \textcolor{black}{42.4} & \textcolor{gray}{0.8} & \textcolor{black}{13.7} & \textbf{\textcolor{black}{85.5}} & \textcolor{gray}{4.5} \\
RequireJS & \textcolor{gray}{2.6} & \textcolor{black}{46.6} & \textbf{\textcolor{black}{50.8}} & \textcolor{black}{8.4} &  & \textcolor{gray}{3.9} & \textbf{\textcolor{black}{95.7}} & \textcolor{gray}{6.4} \\
Angular & \textcolor{gray}{2.2} & \textbf{\textcolor{black}{56.8}} & \textcolor{black}{41.0} & \textcolor{gray}{3.8} &  & \textcolor{gray}{5.7} & \textbf{\textcolor{black}{94.2}} & \textcolor{gray}{5.3} \\
LoDash & \textcolor{gray}{0.7} & \textcolor{black}{48.3} & \textbf{\textcolor{black}{51.0}} & \textcolor{black}{10.6} &  & \textcolor{black}{7.2} & \textbf{\textcolor{black}{92.7}} & \textcolor{black}{6.7} \\
GreenSock & \textcolor{black}{13.6} & \textbf{\textcolor{black}{46.0}} & \textcolor{black}{40.4} & \textcolor{black}{30.5} & \textcolor{black}{20.3} & \textcolor{black}{22.1} & \textbf{\textcolor{black}{57.6}} & \textcolor{gray}{3.7} \\
\bottomrule
\end{tabular}

\label{tab:libraryInclusions}
\end{table}

Figures~\ref{fig:ditrib_detected_alexa} and
\ref{fig:ditrib_detected_com} show the distribution of detected libraries
with respect to their inclusion type (\ie inline, internal, and external)
in the two crawls. We detect libraries in inline script on 4.1\,\%
of all \textsc{Alexa} sites (and 2.8\,\% of \textsc{Com} sites), which
occurs when library code is copy-pasted into a \texttt{<script>} tag
directly embedded in HTML, or when a script is generated by evaluating a
string. For instance, we observe that a version of Dojo loads additional
components by making an asynchronous request and passing the response to
\texttt{eval()}. We detect internal and external inclusions in 62.0\,\%
and 41.1\,\% of all sites with an overall median of two different libraries
per site in \textsc{Alexa} (\textsc{Com}: internal libraries on 21.4\,\% of
sites, external on 48.7\,\%, with a median of one library per site).

Table~\ref{tab:hostRankingLibraries} contains the most frequent hosts
serving detected JavaScript libraries. Unsurprisingly, JavaScript CDNs
are well represented (six in the Top~10 of \textsc{Alexa} and three
in the Top~10 of \textsc{Com}); they were all used to build our
catalogue of library reference files. Between 13\,\% and 18\,\%
of all non-inline JavaScript library inclusions are loaded from
Google's CDN. Perhaps illustrating the ``long tail'' nature of the
\textsc{Com} crawl, we observe that several web hosting and domain
parking companies are represented in the Top~10 library sources of
\textsc{Com} (note that we did not group subdomains; e.g., there are
multiple other subdomains of \texttt{wsimg.com} below the Top~10).

\begin{figure*}[!ht]
\centering
\begin{minipage}{0.48\textwidth}
    \centering
    \capstart
    \includegraphics[width=1.0\textwidth]{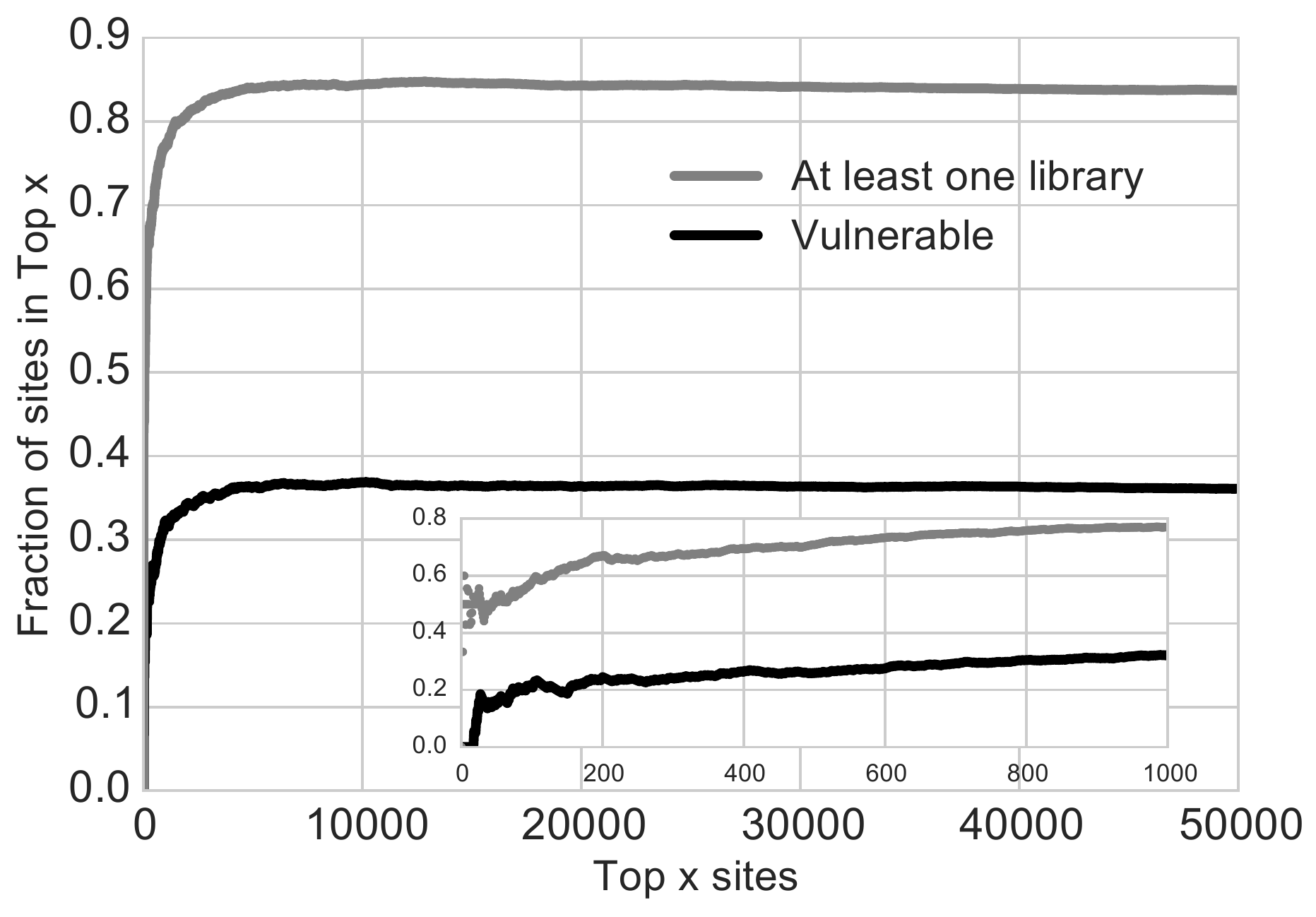}
    \vspace{-0.732cm}
    \caption{Fraction of websites with detected vulnerable inclusions in
             the \textsc{Alexa} Top~$x$, and fraction of websites with at
             least one detected library for comparison. Vulnerable
             libraries can be found on 26\,\% of Top~$500$ websites.}
    \label{fig:distribAffected}
\end{minipage}
\hfill
\begin{minipage}{0.48\textwidth}
    \centering
    \capstart
    \includegraphics[width=0.985\textwidth]{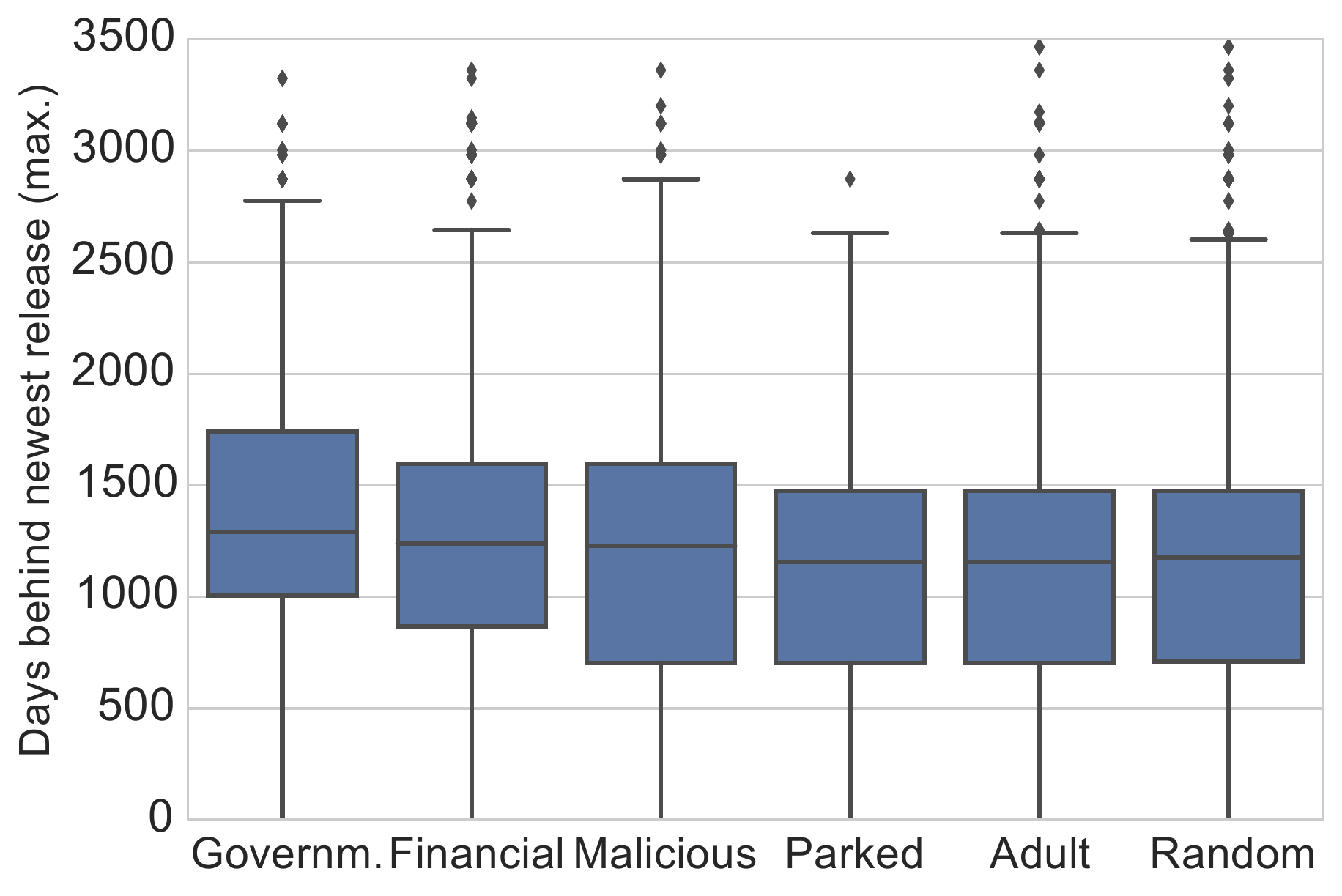}
    \caption{Box plot of each website's maximum lag behind the newest
             release of each detected library, broken down into site
             categories and a random sample of 10,000 websites
             (\textsc{Alexa}). Governmental sites are longest behind.}
    \label{fig:lagCtg}
\end{minipage}
\end{figure*}

Table~\ref{tab:libraryInclusions} lists the percentage of inline,
internal, and external inclusions for a selection of detected libraries.
While the majority of inclusions in \textsc{Alexa} come from
internally-hosted sources (\ie the website's own domain), most
inclusions are external in \textsc{Com}.
About 7.3\,\% of all sites
in \textsc{Alexa} (16.0\,\% in \textsc{Com}) contain at least one
library inclusion that is causally related to \textit{ad},
\textit{tracker} or \textit{widget} code;
SWFObject and GreenSock are relatively often included by such code.

\subsection{Vulnerable Libraries}
\label{sec:analysis:vulnerabilities}
We now move on to answering our research questions, starting with
how many websites become potentially vulnerable due to including a
library version with known vulnerabilities.

Figure~\ref{fig:distrib_vuln_alexa} shows the distribution of total and vulnerable
libraries per \textsc{Alexa} website. Overall, we find that 37.8\,\% of sites use at least one
library version that we know to be vulnerable, and 9.7\,\% use two or more different
vulnerable library versions (\textsc{Com} in
Figure~\ref{fig:distrib_vuln_com}: 37.4\,\% and 4.1\,\%).

To better understand how website popularity is correlated with vulnerable libraries, we plot in
Figure~\ref{fig:distribAffected} the percentage of vulnerable \textsc{Alexa} websites according
to their Alexa ranks. Highly-ranked websites tend to be less likely to
include vulnerable libraries, but they are also less likely to include
any detected library at all. Towards the lower ranks, both curves
increase at a similar pace until they stabilise. While only 21\,\% of the
Top~100 websites use a known vulnerable library, this percentage increases
to 32.2\,\% in the Top~1\,k before it stabilises in the Top~5\,k and
remains around the overall average of 37.8\,\% for all 75\,k websites.
This indicates that even relatively large websites (albeit not the largest
ones) have a vulnerability rate comparable to the average of the
\textsc{Com} dataset, which is dominated by smaller sites.

When grouping \textsc{Alexa} according to the McAfee SmartFilter
web categorisation~\cite{webCategories}, we find that
financial and governmental websites rank {\em last} with 52\,\% and
50\,\% vulnerable sites, respectively. Malicious websites (\eg spam) have
the same proportion as the full dataset, while parked and adult sites
are the least vulnerable with 24\,\% and 19\,\%, respectively.

Table~\ref{tab:vulnerableInclusions} shows the percentage of vulnerable copies
for 5 out of the 11 JavaScript libraries in our catalogue with vulnerability data.
In \textsc{Alexa},
36.7\,\% of jQuery inclusions are known vulnerable, when at most one
inclusion of a specific library version is counted per site. Angular has
40.1\,\% vulnerable inclusions, Handlebars has 86.6\,\%, and YUI 3 has
87.3\,\% (it is not maintained any more). These numbers illustrate that
inclusions of known vulnerable versions can make up even a majority of
all inclusions of a library. However, we caution that these numbers
are not suitable for comparing the relative vulnerabilities of different
libraries (see Section~\ref{sec:libcatalogue:limits}).

\subsection{Risk Factors for Vulnerability}
\label{sec:analysis:factors}

So far, we have examined whether sites are potentially vulnerable, that
is, whether they include one or more known vulnerable libraries. In the
following, we turn to \textit{how} libraries are included by sites and
whether we can identify any factors indicating a higher fraction of
vulnerable inclusions.

When we look at how the library is being hosted,
inline inclusions of jQuery have a clearly higher fraction of vulnerable
versions than internally or externally hosted copies. The situation is
similar when focusing on the parent node of a library, that is, the
document or script that caused the library to be included. For most
libraries, direct inclusions by the main document are less
likely to be vulnerable than indirect inclusions such as through an
intermediate script or in a frame.

As an example for web applications, we briefly survey library use related to
WordPress. We consider a library to be used by WordPress if the URL of
either the library or its next parent in the causality tree contains
\texttt{/wp-content/}; an inclusion is unrelated if neither
condition holds. For all libraries, WordPress-related inclusions appear
to be slightly more up-to-date than unrelated inclusions.

On the other
hand, library inclusions by ad, widget or tracker code appear to be
more vulnerable than unrelated inclusions. While the difference is
relatively small for jQuery in \textsc{Alexa}, the
vulnerability rate of jQuery associated with ad, widget or tracker code
in \textsc{Com}---89\,\%---is almost double the rate of unrelated
inclusions. We speculate that this may
be due to the use of less reputable ad networks or widgets on the
smaller sites in \textsc{Com} as opposed to the larger sites in
\textsc{Alexa}.

We note that the factors above are relatively coarse and the effects are
sometimes opposite for different libraries. We suggest further
investigation into the underlying causes.

\begin{table*}[t]
\caption{Vulnerable fraction of inclusions per library, counting at
         most one library-version pair per site. Configurations with
         less than 100 vulnerable inclusions grayed out; omitted if
         less than 10 total inclusions after filter.}
\centering

\subtable[\textsc{Alexa}]{
\scriptsize

\begin{tabular}{l@{\hspace{0.35cm}}r@{\hspace{0.6cm}}r@{\hspace{0.4cm}}r@{\hspace{0.1cm}}r@{\hspace{0.6cm}}r}
\toprule
\textbf{Inclusion Filter} & \textbf{jQuery} & \textbf{jQ-UI} & \textbf{Angular} & \textbf{Handlebars} & \textbf{YUI 3} \\
\midrule
All Inclusions & \textcolor{black}{36.7\,\%} & \textcolor{black}{33.7\,\%} & \textcolor{black}{40.1\,\%} & \textcolor{black}{86.6\,\%} & \textcolor{black}{87.3\,\%} \\
\addlinespace
Internal & \textcolor{black}{38.1\,\%} & \textcolor{black}{33.0\,\%} & \textcolor{black}{37.1\,\%} & \textcolor{black}{84.6\,\%} & \textcolor{black}{92.8\,\%} \\
External & \textcolor{black}{34.8\,\%} & \textcolor{black}{35.5\,\%} & \textcolor{black}{48.2\,\%} & \textcolor{black}{88.6\,\%} & \textcolor{black}{83.8\,\%} \\
Inline & \textcolor{black}{54.8\,\%} & \textcolor{gray}{30.7\,\%} & \textcolor{gray}{20.0\,\%} & \textcolor{gray}{100.0\,\%} & \textcolor{gray}{-} \\
\addlinespace
Internal Parent & \textcolor{black}{37.1\,\%} & \textcolor{black}{33.7\,\%} & \textcolor{black}{40.6\,\%} & \textcolor{black}{85.3\,\%} & \textcolor{black}{91.3\,\%} \\
External Parent & \textcolor{black}{32.6\,\%} & \textcolor{black}{33.8\,\%} & \textcolor{gray}{41.8\,\%} & \textcolor{black}{96.4\,\%} & \textcolor{black}{92.2\,\%} \\
Inline Parent & \textcolor{black}{47.6\,\%} & \textcolor{gray}{35.2\,\%} & \textcolor{gray}{25.6\,\%} & \textcolor{gray}{83.7\,\%} & \textcolor{black}{79.6\,\%} \\
\addlinespace
Direct Incl.\ in Root & \textcolor{black}{36.4\,\%} & \textcolor{black}{33.5\,\%} & \textcolor{black}{40.1\,\%} & \textcolor{black}{85.2\,\%} & \textcolor{black}{90.4\,\%} \\
Indirect Inclusion & \textcolor{black}{41.0\,\%} & \textcolor{black}{35.6\,\%} & \textcolor{black}{43.2\,\%} & \textcolor{black}{92.3\,\%} & \textcolor{black}{87.1\,\%} \\
\addlinespace
WordPress & \textcolor{black}{29.2\,\%} & \textcolor{gray}{14.4\,\%} & \textcolor{gray}{23.5\,\%} & \textcolor{gray}{77.3\,\%} & \textcolor{gray}{-} \\
Non-WordPress & \textcolor{black}{36.8\,\%} & \textcolor{black}{34.2\,\%} & \textcolor{black}{40.7\,\%} & \textcolor{black}{86.8\,\%} & \textcolor{black}{87.3\,\%} \\
\addlinespace
Ad/Widget/Tracker & \textcolor{black}{38.1\,\%} & \textcolor{gray}{39.8\,\%} & \textcolor{gray}{23.5\,\%} & \textcolor{gray}{96.9\,\%} & \textcolor{black}{98.2\,\%} \\
No Ad/Widget/Tracker & \textcolor{black}{36.7\,\%} & \textcolor{black}{33.6\,\%} & \textcolor{black}{40.8\,\%} & \textcolor{black}{86.5\,\%} & \textcolor{black}{84.1\,\%} \\
\bottomrule
\end{tabular}

	\label{tab:vulnerableInclusions:Alexa}
}
\hfill
\subtable[\textsc{Com}]{
\scriptsize

\begin{tabular}{l@{\hspace{0.35cm}}r@{\hspace{0.6cm}}r@{\hspace{0.4cm}}r@{\hspace{0.1cm}}r@{\hspace{0.6cm}}r}
\toprule
\textbf{Inclusion Filter} & \textbf{jQuery} & \textbf{jQ-UI} & \textbf{Angular} & \textbf{Handlebars} & \textbf{YUI 3} \\
\midrule
All Inclusions & \textcolor{black}{55.4\,\%} & \textcolor{black}{37.3\,\%} & \textcolor{black}{38.7\,\%} & \textcolor{black}{87.5\,\%} & \textcolor{black}{13.7\,\%} \\
\addlinespace
Internal & \textcolor{black}{41.6\,\%} & \textcolor{black}{28.1\,\%} & \textcolor{gray}{45.8\,\%} & \textcolor{gray}{100.0\,\%} & \textcolor{gray}{68.8\,\%} \\
External & \textcolor{black}{62.7\,\%} & \textcolor{black}{42.7\,\%} & \textcolor{black}{38.4\,\%} & \textcolor{black}{85.4\,\%} & \textcolor{black}{12.6\,\%} \\
Inline & \textcolor{black}{89.9\,\%} & \textcolor{gray}{25.6\,\%} & \textcolor{gray}{-} & \textcolor{gray}{-} & \textcolor{gray}{-} \\
\addlinespace
Internal Parent & \textcolor{black}{59.7\,\%} & \textcolor{black}{37.0\,\%} & \textcolor{black}{78.2\,\%} & \textcolor{gray}{94.2\,\%} & \textcolor{gray}{47.9\,\%} \\
External Parent & \textcolor{black}{45.9\,\%} & \textcolor{black}{38.0\,\%} & \textcolor{black}{20.7\,\%} & \textcolor{black}{84.1\,\%} & \textcolor{gray}{68.9\,\%} \\
Inline Parent & \textcolor{black}{79.8\,\%} & \textcolor{gray}{48.9\,\%} & \textcolor{gray}{-} & \textcolor{gray}{-} & \textcolor{gray}{1.0\,\%} \\
\addlinespace
Direct Incl.\ in Root & \textcolor{black}{42.6\,\%} & \textcolor{black}{36.2\,\%} & \textcolor{black}{41.3\,\%} & \textcolor{black}{88.7\,\%} & \textcolor{gray}{50.4\,\%} \\
Indirect Inclusion & \textcolor{black}{77.5\,\%} & \textcolor{black}{44.5\,\%} & \textcolor{gray}{30.8\,\%} & \textcolor{gray}{86.4\,\%} & \textcolor{gray}{7.5\,\%} \\
\addlinespace
WordPress & \textcolor{black}{41.6\,\%} & \textcolor{gray}{20.4\,\%} & \textcolor{gray}{25.0\,\%} & \textcolor{gray}{70.3\,\%} & \textcolor{gray}{-} \\
Non-WordPress & \textcolor{black}{55.6\,\%} & \textcolor{black}{38.6\,\%} & \textcolor{black}{38.9\,\%} & \textcolor{black}{90.7\,\%} & \textcolor{black}{13.7\,\%} \\
\addlinespace
Ad/Widget/Tracker & \textcolor{black}{89.0\,\%} & \textcolor{gray}{31.6\,\%} & \textcolor{gray}{19.2\,\%} & \textcolor{gray}{-} & \textcolor{gray}{55.0\,\%} \\
No Ad/Widget/Tracker & \textcolor{black}{45.5\,\%} & \textcolor{black}{37.3\,\%} & \textcolor{black}{39.6\,\%} & \textcolor{black}{87.3\,\%} & \textcolor{black}{12.7\,\%} \\
\bottomrule
\end{tabular}

	\label{tab:vulnerableInclusions:Com}
}
\label{tab:vulnerableInclusions}
\end{table*}

\subsection{Relative Age of Libraries}
\label{sec:analysis:age}

When websites include outdated libraries, an interesting question is how
far they are behind more current versions of the libraries.
We begin by looking at how far sites are behind the most recent
\textit{patch}-level releases of libraries. Patch-level releases are
usually (though not always) backwards-compatible. In \textsc{Alexa},
we observe that non-vulnerable
inclusions of Angular lag behind by a median of five versions, whereas the median is
seven versions for vulnerable inclusions.

When looking at the number of
days between the release date of the included version and the release date
of the newest available library version overall, the lag for Angular is
398~days for all inclusions, 657~days for vulnerable inclusions, and
234~days for non-vulnerable inclusions. While this may suggest that
there is a relationship between higher lag and vulnerability of a site,
we note that the lag is also tied to the availability of newer
versions---if no update is available, the lag is zero yet the used
version could be vulnerable.
For instance, 6.7\,\% of all (16.6\,\% of vulnerable) Angular inclusions
in \textsc{Alexa} use the latest available release in their respective
\textit{patch} branch, but remain vulnerable as the branch contains no
fixed version.

\begin{table}[t]
\caption{Days between release of used version and newest available
library version (median over inclusions). Shown separately
for vulnerable and non-vulnerable inclusions where available, or
(centred) for all inclusions otherwise. Greyed out when less than
100~inclusions observed.}
\centering 
\scriptsize

\begin{tabular}{lrrrr}
\toprule
 & \multicolumn{2}{c}{\textbf{\textsc{Alexa}}} & \multicolumn{2}{c}{\textbf{\textsc{Com}}} \\
\cmidrule(lr){2-3} \cmidrule(lr){4-5}
\textbf{Library} & \textbf{Vuln.} & \textbf{Non-V.} & \textbf{Vuln.} & \textbf{Non-V.} \\
\midrule
jQuery & \textcolor{black}{1476} & \textcolor{black}{705} & \textcolor{black}{2243} & \textcolor{black}{705} \\
jQuery-Migrate & \textcolor{black}{1105} & \textcolor{black}{1024} & \textcolor{gray}{1105} & \textcolor{black}{1024} \\
jQuery-Mobile & \multicolumn{2}{c}{\textcolor{black}{49}} & \multicolumn{2}{c}{\textcolor{gray}{122}} \\
Bootstrap & \multicolumn{2}{c}{\textcolor{black}{304}} & \multicolumn{2}{c}{\textcolor{black}{389}} \\
Angular & \textcolor{black}{657} & \textcolor{black}{234} & \textcolor{black}{489} & \textcolor{black}{287} \\
Handlebars & \textcolor{black}{687} & \textcolor{black}{0} & \textcolor{black}{687} & \textcolor{gray}{57} \\
Prototype & \multicolumn{2}{c}{\textcolor{black}{1772}} & \multicolumn{2}{c}{\textcolor{black}{1772}} \\
MooTools & \multicolumn{2}{c}{\textcolor{black}{1417}} & \multicolumn{2}{c}{\textcolor{black}{1417}} \\
YUI 3 & \textcolor{black}{393} & \textcolor{gray}{140} & \textcolor{gray}{393} & \textcolor{black}{140} \\
Raphael & \multicolumn{2}{c}{\textcolor{black}{1491}} & \multicolumn{2}{c}{\textcolor{gray}{1491}} \\
D3 & \multicolumn{2}{c}{\textcolor{black}{491}} & \multicolumn{2}{c}{\textcolor{gray}{139}} \\
\bottomrule
\end{tabular}

\label{tab:relativeAge}
\end{table}

With this caveat in mind, Table~\ref{tab:relativeAge} shows the median
lag (in days)
behind the most recently released version of each library.
We observe significant variations. For instance, the median D3 inclusion
in \textsc{Alexa} uses a version 491~days older than the newest D3
release. For Mootools, the lag is 1,417~days.

To characterise lag from a per-site point of view, we calculate the
maximum lag of all inclusions on each site and find that 61.4\,\% of
\textsc{Alexa}
sites are at least one \textit{patch} version behind on one of their
included libraries (\textsc{Com}: 46.2\,\%). Similarly, the median
\textsc{Alexa} site uses a version released 1,177~days
(\textsc{Com}: 1,476~days) before the newest available release of
the library.

Finally, we plot this lag according to the category of \textsc{Alexa}
sites in Figure~\ref{fig:lagCtg}.
Each candlestick shows the 5$^{th}$, 25$^{th}$, 50$^{th}$, 75$^{th}$, and 95$^{th}$
percentiles, respectively. We observe that both governmental and financial websites
are the longest behind current releases with a median lag of 1,293 and
1,239~days, respectively.
Similar to per-category vulnerability rates, parked and adult websites
exhibit values better than the average.

In summary, these results demonstrate that the majority of web developers
are working with library versions released a long time ago.
We observe median lags measured in years, suggesting that web developers
rarely update their library dependencies once they have deployed a site.

\subsection{Duplicate Inclusions}
\label{sec:analysis:duplicates}

\begin{figure*}[t]
\centering
\begin{minipage}{0.48\textwidth}
    \centering
    \capstart
    \includegraphics[trim={110 295 105 35},clip,width=1.0\textwidth]{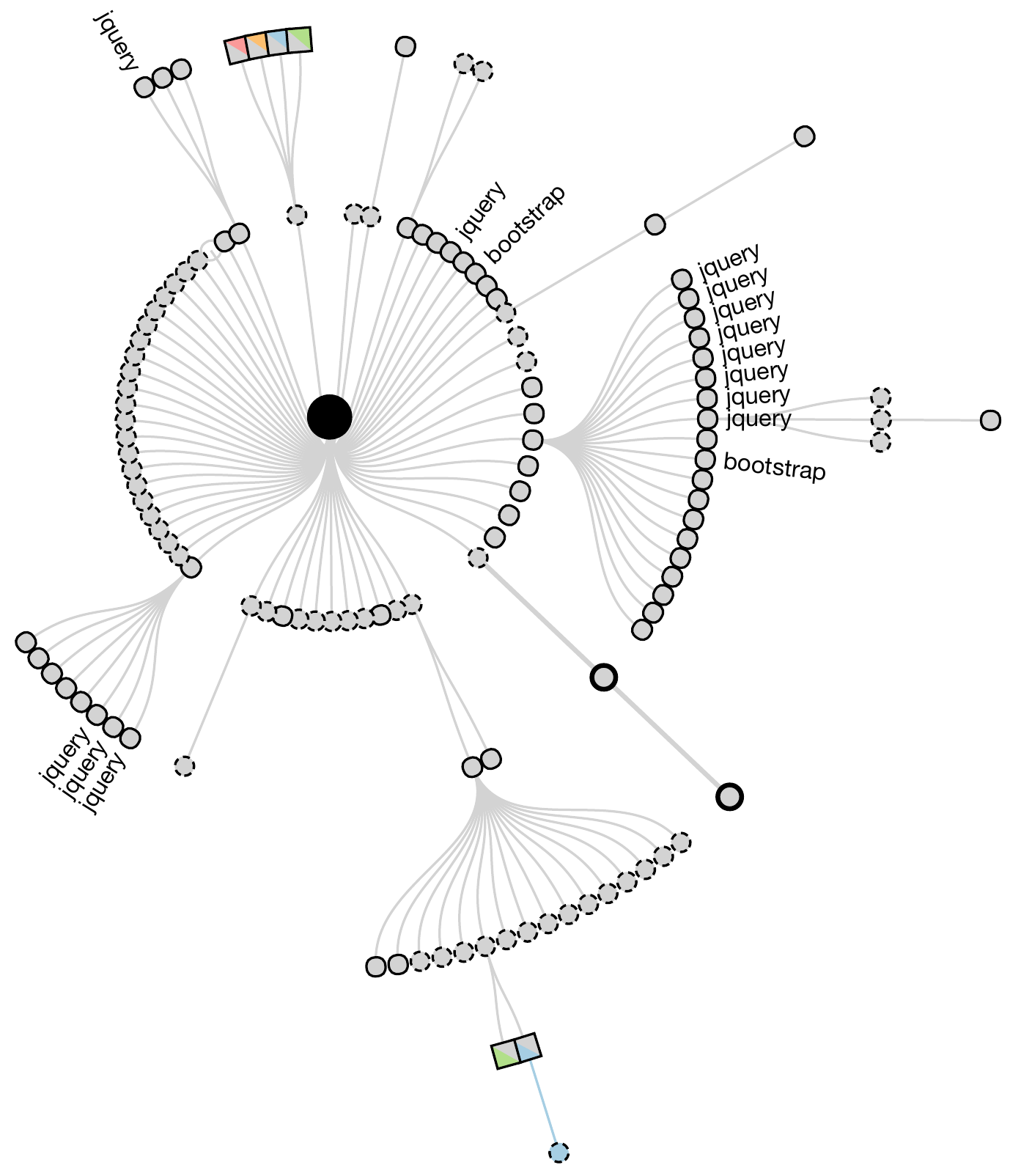}
    \vspace{0.3cm}
    \caption{Causality tree for \texttt{ms.gov}, showing multiple jQuery
             inclusions in a single document. One instance is referenced
             in the main HTML page; all others are included transitively
             by other scripts.}
    \label{fig:ms}
\end{minipage}
\hfill
\begin{minipage}{0.48\textwidth}
    \centering
    \capstart
    \includegraphics[clip,width=1.0\textwidth]{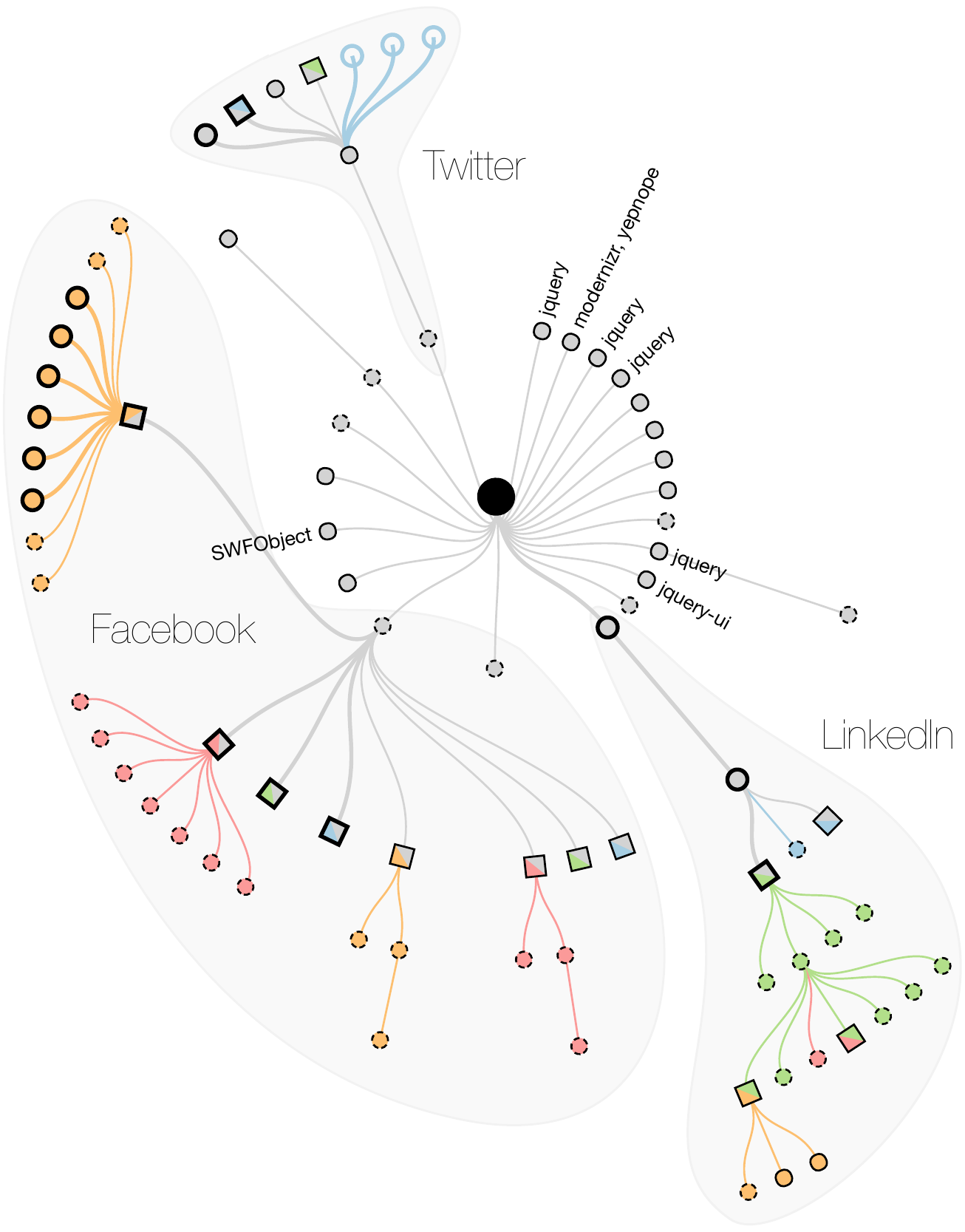} 
    \vspace{-0.7cm}
    \caption{Causality tree for \texttt{mercantil.com} with multiple
             jQuery inclusions, two libraries concatenated into one file
             (Modernizr and Yepnope), and complex element relationships
             in social media widgets (Twitter, Facebook and
             LinkedIn).}
    \label{fig:mercantile}
\end{minipage}
\end{figure*}

While analysing the use of JavaScript libraries on websites, we noticed
that libraries are often used in unexpected ways. We discuss some examples
using jQuery as a case study. About 20.7\,\% of the websites including
jQuery in \textsc{Alexa} (17.2\,\% in \textsc{Com}) do so two or more
times. While it may be necessary to include a library multiple times
within different documents from different origins, 4.2\,\% of websites
using jQuery in \textsc{Alexa} include the same version of
the library two or more times {\em into the same document}
(5.1\,\% in \textsc{Com}), and 10.9\,\% (5.7\,\%)
include two or more \textit{different} versions of jQuery into the same document.
Since jQuery registers itself as a \texttt{window}-global variable, unless
special steps are taken only the last loaded and executed instance can
be used by client code. For asynchronously included instances, it may even
be difficult to predict which version will prevail in the end.

Figure~\ref{fig:ms} shows the causality tree of \texttt{ms.gov}, the
site with the highest number of identical jQuery inclusions in a single
document. Only one instance
(version \texttt{2.2.2}) is included directly in the source code of the
main HTML page; all twelve other jQuery inclusions (of version
\texttt{2.2.0}) are injected by various self-hosted scripts in quick
succession.

In contrast, the inclusions of four \textit{different}
jQuery versions on \texttt{mercantil.com} (Figure~\ref{fig:mercantile})
are all referenced directly in the main page's source code, some of them
directly adjacent to each other. While we can only speculate why these
cases occur, at least some of them may
be related to server-side templating, or the combination of
independently developed components into a single document. Indeed, we
have observed cases where a web application (\eg a WordPress plug-in)
that bundled its own version of a library was integrated into a page
that already contained a separate copy of the same library. Since
duplicate inclusions of a library do not necessarily break any
functionality, we suspect that many web developers may not be aware that
they include a library multiple times, and even less that the duplicate
inclusion may be potentially vulnerable.

\subsection{Remediations}
\label{sec:analysis:aliasing}

From a remediation perspective, the picture painted by our data is bleak.
We observe that only very small fraction of potentially vulnerable sites (2.8\,\%
in \textsc{Alexa}, 1.6\,\% in \textsc{Com}) could become
free of vulnerabilities by applying \textit{patch}-level updates, \ie an
update of the least significant version component, such as from
\texttt{1.2.3} to \texttt{1.2.4}, which would generally be expected to be
backwards compatible. The vast majority of sites would need to install at least one
library with a more recent \texttt{major} or \texttt{minor} version, which
might necessitate additional code changes due to incompatibilities.

\para{Version Aliasing.} Some JavaScript CDNs support
\textit{version aliasing}, where the
developer may specify only a prefix of the requested library version and
the CDN will automatically return the latest available version with that
prefix. In theory, version aliasing appears to be a robust strategy for
developers to easily keep their library dependencies up-to-date.
We scan our crawl for library inclusions with URLs of a CDN,
and detect version aliasing whenever (1) the version given in the URL has
only one or two components, such as \texttt{1.2}, and (2) the library
version detected by the static or dynamic method is greater than the
prefix extended with zeros, such as
\texttt{1.2.3} instead of \texttt{1.2.0}.
In \textsc{Alexa} (and \textsc{Com}), we detect 1,489 (914) confirmed instances
of version aliasing, counting at most one per library on each
site. Overall, however, the frequency of version
aliasing is very small---only around 1.2\,\% of all sites that include jQuery use version
aliasing.

Except for one, all instances of aliasing refer to Google's CDN, with jQuery being the
most frequent library. One one hand, 47.2\,\% (37.9\,\%) of jQuery inclusions
with aliasing are avoiding a vulnerability, \ie the inclusions point to a branch
that has a known vulnerability, but the issue is addressed by the latest version.
On the other hand, while version aliasing may seem like a good way to automatically
avoid vulnerabilities, Google recently discontinued this
service, citing caching issues and ``lack of compatibility between even
minor
versions''~\cite{googleCDN}.

\section{Discussion}
\label{sec:discussion}
Our research has shown that even though patches may be available,
vulnerable JavaScript libraries are in widespread use on the Web. In the
following, we discuss approaches that we believe could improve the
situation.

\subsubsection*{Dependency Management}
Before website developers can update potentially vulnerable libraries
that they are using, they must be aware of which libraries they are using.
Instead of manually copying library files or CDN links into their
codebase, developers should consider more systematic approaches to
dependency management. Bower~\cite{bower} and the more server-oriented
Node Package Manager~\cite{npm} allow developers to declare external
dependencies in a configuration file and can automatically download and
include the code into the project. Tools such as Auditjs~\cite{auditjs}
(for Node projects) scan dependencies for known vulnerabilities and can be
integrated into automated build processes; such solutions, however,
work only if the developer has an understanding of the risks associated
with using vulnerable libraries, and is aware of the audit tool itself.
Therefore, this functionality would ideally be integrated into the
dependency management system of the programming platform so that a
warning can be shown each time a developer includes a known
vulnerable component from the central repository.

\subsubsection*{Code Maintenance}
Effective strategies to have web developers update vulnerable libraries
work only when vulnerability information is properly tracked and
disseminated. Unfortunately, security does not appear to be a priority
in the JavaScript library ecosystem. Popular vulnerability databases
contain nearly no entries regarding JavaScript libraries.
None of the 12 most popular libraries from Table~\ref{tab:libraries}
had a dedicated mailing list for security announcements.
Furthermore, only a few JavaScript library
developers provide a dedicated email address where users can submit
vulnerability reports.
When the release notes of libraries mention at all that a vulnerability
has been fixed, they often do not provide any details about the affected
code, or which prior versions are vulnerable. This is problematic
because web developers using that library do not know whether the
vulnerable code is a function that they depend on, and whether an update
is required. Libraries can even silently reintroduce vulnerabilities in
order to remain backwards-compatible~\cite{heiderich13}. Although jQuery
is an immensely popular library, the fact that searching for ``security''
or ``vulnerability'' in the official learning centre returns
{\em ``Apologies, but nothing matched your search criteria''} is an
excellent summary of the state of JavaScript library security on the
Internet, circa August 2016.\footnote{Ember, the 50\textsuperscript{th} most popular
library in \textsc{Alexa} (rank 52 in \textsc{Com}), is a notable
exception with long-term support versions, a security mailing list, CVEs
for vulnerabilities, and affected versions listed in security notices.}
A similar lack of adequate information about
security issues has also been reported for the Android library
ecosystem~\cite{backes16}.

An additional complication is that patches are often supplied only for
the most recent versions of a library. Yet, these versions are not
necessarily backwards-compatible with the versions still in use by many
web developers. In fact, the short lifecycles common in web development
can become a burden for developers who need to keep up with frequent
breaking API changes to maintain their websites free of vulnerable
libraries.

\subsubsection*{Third-Party Components}
We observed that libraries included by third-party components such as
advertising, tracking or social media widget code have a higher rate
of vulnerability than other inclusions. Such components are often hosted
on third-party servers and loaded dynamically through client-side
JavaScript. Additional libraries loaded at runtime by these components
do not appear in the website's codebase, and web developers may be unaware
that they are indirectly including vulnerable code into their website.
Similarly, dynamic inclusions of libraries by third-party components
may explain some of the same-document duplicate inclusions that we
noticed.
In addition to keeping their library dependencies up to date, developers
of web services meant to be included into other websites could avoid
replacing an existing library instance by using a methodology similar to
ours to test whether the library has already been loaded into the page
before adding their own copy.
On the other hand, web developers who intend to use third-party
components such as advertising code for their website can attempt to limit
potential damage by isolating these components in separate frames
whenever this is feasible.

\section{Related Work}
Our work is related to prior studies on JavaScript security,
measurements of vulnerability patching and dependency management, a
series of blog posts about inclusions of vulnerable libraries that
inspired our more in-depth analysis, and existing tools that implement
a subset of our detection methodology.

\subsubsection*{JavaScript Security}

In~\cite{nikiforakis-2012-ccs}, Nikiforakis \etal identify the network
sources of JavaScript inclusions in the Alexa Top~10\,k websites, without
special consideration for libraries or corresponding versioning semantics,
and develop host-based metrics for maintenance quality to assess whether
remote code providers could be compromised by attackers and subsequently
serve malicious JavaScript. Whereas Nikiforakis \etal study where
included code is hosted, we focus on the narrower but semantically
richer setting of libraries to investigate whether included code is
outdated or known to be vulnerable, and we leverage our deep browser
instrumentation to determine the initiators and causes of such inclusions.

A separate class of related work examines specific attack vectors in
client-side JavaScript and conducts crawls to estimate how many websites
are subject to the attack: Lekies and Johns~\cite{lekies12} survey
insecure usage of JavaScript's \texttt{localStorage()} function for code
caching purposes, Son and Shmatikov~\cite{son13} examine vulnerabilities
arising from unsafe uses of the \texttt{postMessage()} function, Lekies
\etal~\cite{johns13ccs} detect and validate DOM-based XSS vulnerabilities,
and Richards \etal~\cite{richards11} analyse websites' usage patterns of
the problematic \texttt{eval()} API. Yue and Wang~\cite{yuewang13} study
several insecure practices related to JavaScript, namely cross-domain
inclusion of scripts as well as the execution and rendering of dynamically
generated JavaScript and HTML through \texttt{eval()} and
\texttt{document.write()}, respectively. Li \etal~\cite{redfox} detect
malicious redirection code hidden in JavaScript files on compromised
hosts by deriving signatures from the differences between infected library
files and the original, benign copies. In contrast to the above work, we
do not focus on specific vulnerabilities, the use of security-critical
functions, or malicious files. Instead, we provide empirical results at
a more abstract level to highlight and explain the prevalence of
benign-but-vulnerable JavaScript libraries in the wild.

\subsubsection*{Vulnerability and Dependency Management}

Four studies have examined vulnerability patching and dependency management
in large software ecosystems, although not with respect to JavaScript or the Web.
Sonatype Inc., the company behind Maven, released a report~\cite{sonatype}
examining security maintenance practices observed from the vantage point
of the largest repository of Java components. According to the report,
the mean time-to-repair of a security vulnerability in component
dependencies is 390~days, 51,000 of the components in the repository have
known security concerns, and 6.2\,\% of downloaded
components include known vulnerabilities. A key observation in the
report is that fixing serious flaws in open source code does not stop
vulnerable versions from being used. Our
work in this paper shows that there are similar trends with respect to
JavaScript library usage on the Web.

Nappa \etal analyse the patch deployment process for 1,593
vulnerabilities in 10 applications installed on 8.4~million Windows hosts
worldwide~\cite{nappa15}. The authors show that the time until a patch is
released for different applications affected by the same vulnerability
in a shared library can differ by up to 118~days, with a median
of 11~days. Furthermore, patching rates vary among applications and
depend, among other factors, on the update mechanism. At most 14\,\% of
vulnerable hosts are patched before an exploit is released.

Thomas \etal propose an exponential decay model to estimate patching
delays of Android devices~\cite{Thomas2015}. According to the model,
when a new version of the operating system is released, it takes
3.4~years to reach 95\,\% of the devices.

Backes \etal build \textsc{LibScout}~\cite{backes16}, a system to
detect third-party library code in Android applications using a static
approach based on abstracted package trees and method signatures. They
find that 70.4\,\% of library inclusions in their dataset include
an outdated version, and it takes developers an average of almost one year
to migrate their applications to a newer library version after the library
has been updated. In a case study of two vulnerabilities, the
authors show that the average update delay is 59
and 188~days after the library patch is first made available, while some
applications remain without any update. Furthermore, 10 out of 39
advertising libraries contain one or more versions that improperly use
cryptographic APIs.
In contrast to \textsc{LibScout}, our detection approach requires that the
library API methods used in our signatures not be renamed or removed.
While a theoretical possibility, we believe that such eager minification
settings are exceedingly rare on the Web since they would necessitate
processing all code potentially referencing the library, including in
HTML attributes and inline script. To the best of our knowledge, the
default settings of minifiers typically do not rename methods or
remove dead code in client-side JavaScript (see, for instance, the Closure
Compiler~\cite{closureCompilerMinificationSettings}). This assumption
allows us to detect the version of a JavaScript library more reliably
since most libraries self-identify via their \texttt{version} attribute
or method.

\subsubsection*{Blog Posts}

In 2014, a series of blog posts by Oftedal~(\cite{blog1,blog2,blog3})
raised awareness about the use of outdated
JavaScript libraries on the Web, and the fact that many large companies
(including banks) use versions that are known to be
vulnerable. We complement this first exploration of the issue with a more
comprehensive detection methodology and a more detailed analysis. To the
best of our knowledge, we are the first to report on the modality and
causes of JavaScript library inclusions in websites, uncovering issues
such as duplicate library inclusions as well as transitive (and on
average more vulnerable) inclusions of libraries by third-party modules
such as advertising, tracking, and social media widget code.

\subsubsection*{Tools}
From the point of view of our library detection methodology, we are aware
of two open source tools with a similar approach:
Retire.js and the Library Detector extension.

\noindent\textbf{Library Detector Chrome Extension.} This browser
extension~\cite{librarydetector} aims to detect the JavaScript libraries
running on a website. It injects a script into the website's main document
to test for the presence of known libraries and extracts their version,
using dynamic detection code similar to the approach presented in
Section~\ref{sec:identification}. The extension does not warn against
known vulnerabilities, does not reveal how or why a library was included,
cannot reliably detect duplicate inclusions, and does not analyse
libraries loaded in frames.

\noindent\textbf{Retire.js.} Along with his blog posts, Oftedal released
a tool~\cite{retirejs} to help web developers detect JavaScript libraries
with known vulnerabilities.
In a nutshell, Retire.js is a browser extension that intercepts network requests for
JavaScript files while a website is loading and detects libraries based on
known file hashes, regular expressions over the file contents, and API
method signatures dynamically evaluated in an empty sandbox
environment.
While we also use dynamic detection and hash detection approaches in our
methodology, Retire.js makes several simplifications that limit the tool's
utility for our analysis. First, detecting a script as a library in an
empty sandbox fails when the library has unmet dependencies. jQuery-UI,
for instance, requires jQuery and hence cannot be detected dynamically if
jQuery is not present in the environment. Second, intercepting requests
only at the network level may miss inline scripts, dynamically evaluated
scripts, and duplicate inclusions of cached scripts. Most importantly,
Retire.js does not reveal why a library was included, that is, whether
the inclusion was caused by advertising code, for instance.
We support all of these scenarios and found interesting
results as a consequence, such as the vulnerability rates per inclusion
type in Table~\ref{tab:vulnerableInclusions}, and the duplicate
inclusions observed in Section~\ref{sec:analysis:duplicates}.

\section{Conclusion}
\label{sec:conclusion}

Third-party JavaScript libraries such as Angular, Bootstrap and jQuery
are are frequently used on websites today. While such libraries allow web
developers to create highly interactive, visually appealing
websites, vulnerabilities in these libraries might increase the attack
surface of the websites that depend on them. Hence, it is very important
to ensure that only recent, patched versions of these libraries are being
utilised.

In this paper, we presented the first comprehensive study on the security
implications surrounding JavaScript library usage on real-world websites.
We found that:
\begin{packed_itemize}
\item 86.6\,\% of \textsc{Alexa} Top~75\,k websites and 65.4\,\% of
\textsc{Com} websites use at least one of the 72 JavaScript libraries
in our catalogue (Section~\ref{sec:analysis:detections});

\item more than 37\,\% of websites use at least one library version with
a known vulnerability, and vulnerable inclusions can account for a
significant portion of all observed inclusions of a library
(Section~\ref{sec:analysis:vulnerabilities});

\item the median lag between the oldest library version used on each
website and the newest available version of that library is 1,177~days in
\textsc{Alexa} and 1,476~days in \textsc{Com}
(Section~\ref{sec:analysis:age}), and development of some libraries still
in active use ceased years ago;

\item surprisingly often, libraries are not referenced directly in a page,
but also inlined, or included transitively by other content such as
advertising, tracking or social media widget code
(Table~\ref{tab:libraryInclusions}), and those inclusions have a higher
rate of vulnerability than other, direct inclusions
(Table~\ref{tab:vulnerableInclusions},
Section~\ref{sec:analysis:factors});

\item composition of content modules or third-party content in the same
document can lead to duplicate inclusions of a library and potentially
nondeterministic behaviour with respect to vulnerability
(Section~\ref{sec:analysis:duplicates});

\item remediation efforts are hindered by a lack of backwards-compatible
patches (Section~\ref{sec:analysis:aliasing}) and, more generally, scant
availability of information (Section~\ref{sec:discussion}).
\end{packed_itemize}

The results of this work highlight the need for more thorough and
systematic approaches to JavaScript library inclusion and
dependency management on the Web.

\vspace{0.3cm}
\noindent\textit{The causality trees shown in this work can be viewed
online: \url{https://seclab.ccs.neu.edu/static/projects/javascript-libraries/}}
\vspace{0.55cm}

\section*{Acknowledgements}
This work was supported by the National Science Foundation under grants
CNS-1409738 and CNS-1563320.
\vspace{0.215cm}

{
\balance
\bibliographystyle{IEEEtranS}
\bibliography{references}
}

\end{document}